\documentclass[twocolumn, twocolappendix]{aastex631}

\shorttitle{YunMa: Enabling Spectral Retrievals of Exoplanetary Clouds}
\shortauthors{Ma et al.}
\graphicspath{{./}{figures/}}
\usepackage{savesym}
\savesymbol{tablenum}
\usepackage{siunitx}
\usepackage{chemfig}
\usepackage{amssymb} 
\usepackage{latexsym}
\usepackage{graphicx}
\usepackage{longtable}
\usepackage{epsf}
\usepackage{color, colortbl}

\definecolor{Gray}{gray}{0.9}
\definecolor{Dark}{gray}{0.7}
\definecolor{LightCyan}{rgb}{0.88,1,1}
\definecolor{LightYellow}{rgb}{1,1,0.88}
\definecolor{LightPink}{rgb}{1,0.88,1}
\definecolor{LightRed}{rgb}{1,0.88,0.88}
\definecolor{LightPurple}{rgb}{0.88,0.88,1}

\begin{document}

\title{YunMa: Enabling Spectral Retrievals of Exoplanetary Clouds}

\correspondingauthor{Sushuang Ma}
\email{sushuang.ma.20@ucl.ac.uk}
\footnote{This is the Accepted Manuscript version of an article accepted for publication in \apj. IOP Publishing Ltd is not responsible for any errors or omissions in this version of the manuscript or any version derived from it. This Accepted Manuscript is published under a CC BY licence. The Version of Record is available online at \url{http://doi.org/10.3847/1538-4357/acf8ca}.}

\author[0000-0001-9010-0539]{Sushuang Ma}
\affiliation{Department of Physics and Astronomy, University College London \\
Gower Street, WC1E 6BT London, United Kingdom}

\author[0000-0002-0598-3021]{Yuichi Ito}
\affiliation{National Astronomical Observatory of Japan \\
2 Chome-21-1 Osawa, Mitaka, Tokyo, 181-8588, Japan}
\affiliation{Department of Physics and Astronomy, University College London \\
Gower Street, WC1E 6BT London, United Kingdom}

\author[0000-0003-2241-5330]{Ahmed Faris Al-Refaie}
\affiliation{Department of Physics and Astronomy, University College London \\
Gower Street, WC1E 6BT London, United Kingdom}

\author[0000-0001-6516-4493]{Quentin Changeat}
\affiliation{European Space Agency (ESA) \\
ESA Office, Space Telescope Science Institute (STScI), 3700 San Martin Drive, Baltimore MD 21218, United States of America}
\affiliation{Department of Physics and Astronomy, University College London \\
Gower Street, WC1E 6BT London, United Kingdom}

\author[0000-0002-5494-3237]{Billy Edwards}
\affiliation{SRON, Netherlands Institute for Space Research, Niels Bohrweg 4, NL-2333 CA, Leiden, The Netherlands\\}
\affiliation{AIM, CEA, CNRS, Université Paris-Saclay, Université de Paris, F-91191 Gif-sur-Yvette, France}
\affiliation{Department of Physics and Astronomy, University College London \\
Gower Street, WC1E 6BT London, United Kingdom}

\author[0000-0001-6058-6654]{Giovanna Tinetti}
\affiliation{Department of Physics and Astronomy, University College London \\
Gower Street, WC1E 6BT London, United Kingdom}



\received{2023 January 19}
\revised{2023 July 14}
\accepted{2023 September 1}
\published{2023 November 3}
\submitjournal{\apj}

\begin{abstract}

In this paper, we present \emph{YunMa}, an exoplanet cloud simulation and retrieval package, which enables the study of cloud microphysics and radiative properties in exoplanetary atmospheres. \emph{YunMa} simulates the vertical distribution and sizes of cloud particles and their corresponding scattering signature in transit spectra. We validated \emph{YunMa} against results from the literature. 

When coupled to the \emph{TauREx 3} platform, an open Bayesian framework for spectral retrievals, \emph{YunMa} enables the retrieval of the cloud properties and parameters from transit spectra of exoplanets. The sedimentation efficiency ($f_{\mathrm{sed}}$), which controls the cloud microphysics, is set as a free parameter in retrievals. We assess the retrieval performances of \emph{YunMa} through 28 instances of a K2-18 b-like atmosphere with different fractions of H$_2$/He and N$_2$, and assuming water clouds. Our results show a substantial improvement in retrieval performances when using \emph{YunMa} instead of a simple opaque cloud model and highlight the need to include cloud radiative transfer and microphysics to interpret the next-generation data for exoplanet atmospheres. This work also inspires instrumental development for future flagships by demonstrating retrieval performances with different data quality. 

\end{abstract}




\section{Introduction} \label{sec:intro}


Thousands of exoplanets have been detected since the late $20^{\mathrm{th}}$ century. During the past decade, transit spectroscopy has become one of the most powerful techniques for studying exoplanets' atmospheres in-depth \citep[e.g., reviews by][]{Tinetti2013aareview, Burrows2014review, madhusudhan2019exoplanetary}. Data recorded from space-borne instruments (e.g., Hubble, Spitzer and James Webb Space Telescopes) or from the ground have revealed important information about exoplanet atmospheric chemistry and dynamics \citep[e.g.,][]{sing2016continuum, Tsiaras2018population, Welbanks2019Trends, changeat2022five, Edwards2022G141, jwsters2022wasp39b, venot2020global, Roudier2021Diseq} and may provide insight into planetary interior composition and formation \citep{madhusudhan2020interior, yu2021identify, tsai2021infer, charnay2022titan}. 

A number of spectral retrieval models have been developed by different teams to interpret the atmospheric data and quantify their information content; these include e.g., \citet{MadhusudhanSeager2009retrieval}, \citet{Lee2012retrieval}, TauREx 3 \citep{alrefaie2021taurex31}, NEMESIS \citep{irwin2008nemesis}, CHIMERA \citep{line2013systematic}, ARCiS \citep{min2020arcis, ormel2019arcis}, PICASO \citep{robbins-blanch2022picaso, Batalha2019picaso}, BART \citep{Harrington2022Bart}, petitRADTRANS \citep{Molli2020petitTRANS}, HELIOS \citep{Kitzmann2020helios}, POSEIDON \citep{macdonald2017hd}, HyDRA \citep{Gandhi2018HyDRA}, SCARLET \citep{Benneke2015Scarlet}, PLATON II \citep{Zhang2020platon}, and Pyrat-Bay \citep{Cubillos2021PYRAT}. Up to date, most of the retrieval studies of exoplanetary atmospheres are highly parameterised. This approach has been very sensible given the relatively poor information content of current 
atmospheric data. However, a number of papers in the literature \citep[e.g.,][]{Caldas2019bias, changeat2021bias, changeat2022five} have cautioned against this approach when applied to data recorded with next-generation facilities.


Clouds are omnipresent in planetary, exoplanetary and brown dwarf atmospheres \citep[see e.g.,  review by][]{helling2022cloud} and have often been detected in exoplanet atmospheric data \citep{kreidberg2014clouds, sing2016continuum, Stevenson2016cloudpopulation, Tsiaras2018population}. Their presence imposes additional complexity and uncertainties in the interpretation of exoplanet atmospheric spectra  \citep[e.g.,][]{changeat2021disentangling, tsiaras2019water, Mai2019cloudcompare}. 

Models simulating the formation and radiative properties of clouds and hazes have been published in the literature, e.g., Exo-REM \citep{baudino2015exorem, charnay2018exorem}, \citet{Gao2020cloudspec}, \citet{Windsor2023cloud} and \citet{kawashima2018theoretical}.

Due to the -- currently limited -- observational constraints and computational resources available to simulate the complexity of clouds, retrieval studies of cloudy atmospheres are still in their infancy \citep[see e.g.,][]{Fortney2021review}. 
For instance, many studies have adopted wavelength-independent opaque clouds, where all the radiation beneath the cloud top is blocked from reaching the telescope, and retrieve the vertical location of clouds \citep{boucher2021gray, brogi2019grey}. \citet{wakeford2018cloudretrieval} used a grey, uniform cloud in the ATMO Retrieval Code (ARC) \citep{Goyal2017ATMO,Drummond2016ATMO,Tremblin2015ATMO}.
Other models constrain from radiative transfer the uniform cloud particle sizes without being estimated through cloud microphysics models. For instance, \citet{benneke2019sub} have initially estimated the particle sizes in the atmosphere of GJ 3470 b using Mie-scattering theory.
Extended from this highly parametric approach, cloud scattering parameters and inhomogeneous coverage were also retrieved: NEMESIS was used by \citet{barstow2020parametric} and \citet{wang2022nongray} to retrieve the cloud's opacity, scattering index, top and base pressures, particle sizes and shape factor. \citet{pinhas2019POSEIDION} run POSEIDON to constrain the cloud's top pressure and coverage fraction. \citet{wang2022nongray} adopted PICASO to extract the cloud's base pressure, optical thickness, single scattering albedo, scattering asymmetry and coverage. \citet{lueber2022cloud} extended the use of Helios-r2 to retrieve non-grey clouds, with extinction efficiencies estimated from Mie theory calculations. The model Aurora \citep{welbanks2021aurora} presents inhomogeneities in cloud and haze distributions by separating the atmosphere horizontally into four distinct areas.

The data provided by the next-generation telescopes will be greatly superior in quality and quantity, allowing us to obtain more stringent constraints to our understanding of clouds in exoplanetary atmospheres. Transit spectra of exoplanets recorded from space by the James Webb Space Telescope \citep[JWST, 0.6--28.3 \si{\micro m},][]{Bean2018jwst,Greene2016jwst,gardner2006james}, Ariel \citep[0.5--7.8 \si{\micro m},][]{tinetti2018chemical,tinetti_ariel2} and Twinkle \citep[0.5--4.5 \si{\micro m},][]{edwards2019exoplanet} at relatively high spectral resolution and/or broad wavelength coverage will open the possibility of integrating self-consistent, cloud microphysics approaches into atmospheric retrieval codes. 
A good example of such models is ARCiS \citep[ARtful modelling Code for exoplanet Science,][]{min2020arcis,ormel2019arcis}, which simulates cloud formation from diffusion processes and parametric coagulation. ARCiS also generates cloudy transit spectra from Mie theory \citep{mie1984scattering} and Distribution of Hollow Spheres \citep[DHS,][]{Min2005DHS,Molli2019DHS}, and can be used to retrieve the cloud diffusivity and nuclei injection from transit spectra.

In this work, we present a new optimised model to study cloud microphysical processes directly integrated into a spectral retrieval framework. 
We consider clouds as a thermochemical product, i.e. the aggregation of condensates in the atmosphere, while hazes form photochemically \citep{kawashima2018theoretical}. 
The cloud distribution depends on the atmospheric conditions. Being generated thermochemically, clouds form and diffuse depending on the atmospheric thermal structure and, in return, contribute to it. They also depend on the mixing profiles of the condensable gases in the atmosphere. Clouds act as absorbers and/or scatterers and therefore may dampen the atomic and molecular spectroscopic features and change the continuum. 

Based on studies of the Earth and Solar System's planetary atmospheres, \citet{lewis1969clouds} published a  1-D cloud model  optimised to describe tropospheric clouds in giant planets.  This model assumes that the fall speeds of all condensates are equivalent to the updraft velocities, and only vapour is transported upward. \citet{lunine1989effect} included a correlation between cloud particle sizes, downward sedimentation and upward turbulent mixing. Based on previous models by \citet{lewis1969clouds}, \citet{carlson1988cloud}, \citet{lunine1989effect} and \citet{marley1999reflected}, \citet{ackerman2001precipitating} proposed a new method to estimate the mixing ratio and vertical size distribution of cloud particles (A-M model hereafter). In the A-M model, the sedimentation timescale is estimated through cloud microphysics, taking into account the atmospheric gas kinetics and dynamical viscosity. The model assumes an equilibrium between upward turbulent mixing and sedimentation, where the turbulent mixing is derived from the eddy diffusion in the atmosphere. The key assumptions of the A-M model are as follows:
\begin{enumerate}
\item Clouds  are distributed uniformly in the horizontal direction.
\item Condensable particles rain out at (super)saturation while maintaining a balance of the upward and downward drafts.
\item It does not consider the cloud cover variations caused by precipitation or the microphysics between different types of clouds.
\end{enumerate}
The A-M model was originally proposed for giant exoplanets and brown dwarfs and was tested on Jupiter's ammonia clouds, demonstrating that this approach is applicable to a broad range of temperatures and planetary types. 

Another popular 1-D cloud microphysics model is the Community Aerosol and Radiation Model for Atmospheres (CARMA), initially developed for the Earth's stratospheric sulfate aerosols \citep{Turco1979CARMA,Toon1979CARMA}. CARMA is a time-dependent cloud microphysics model which solves the discretised continuity equations for aerosol particles starting from nucleation. 
\citet{gao2018sedimentation} extended the use of CARMA to simulate clouds on giant exoplanets and brown dwarfs by including additional condensates predicted to form in hot atmospheres and compared the results with the A-M model. 
The A-M model, while able to provide the cloud particle sizes and number density distributions, is of intermediate numerical complexity and, therefore, potentially adaptable to be included in retrieval codes. 
In addition to the original implementation by \citet{ackerman2001precipitating}, Virga \citep{rooney2022fsed} simulates the cloud's particle size distribution from the A-M approach and estimates separately the sedimentation efficiency. PICASO \citep{robbins-blanch2022picaso, Batalha2019picaso} adopts Virga to simulate cloudy exoplanetary atmospheres. \citet{adams2022hotjupiters} couples MIT GCM and Virga to include clouds in 3-D models.
The above are forward simulations only. One step further, \citet{xuan2022cloudbrown} present retrieval studies on HD 4747 B with clouds using petitRADTRANS \citep{Molli2020petitTRANS}, where the cloud simulation in retrieval is motivated by the A-M approach. To estimate the cloud mixing ratio in the retrieval iterations, this model does not solve the full ODE (see Equation 2). Instead, it adopts an approximation of the A-M approach, assuming the mixing ratio of condensable gas above the cloud base is negligible.

To simulate inhomogeneities for cloud formation in the horizontal direction, we would need to consider global circulation atmospheric effects, such as those modelled in \citet{Cho_2021}. An example of a 3-D atmospheric model with clouds is Aura-3D \citep{welbanks2021aurora, nixon2022aura3d}. The retrieval part for Aura-3D is highly parametrised, both for the atmospheric and cloud parameters. \citet{Helling2023gridcloud, helling2019cloudmap} have simulated global cloud distributions by generating inputs to their kinetic cloud model from pre-calculated 3-D Global Circulation Models (GCMs). Unfortunately, these complex models require excessive computing time. In addition, the data expected to be observed in the near future are unlikely to constrain the large number of parameters needed in a 3-D model. Therefore, while theoretical studies with 3-D models are very important to progress in our understanding of clouds in exoplanetary atmospheres and as benchmarks, they are currently less useful for interpreting available data.

In this paper, we present a new cloud retrieval model, \emph{YunMa}, optimised for transit spectroscopy. In \emph{YunMa}, we built the cloud model based on \citet{ackerman2001precipitating} and simulated the cloud contribution to transit spectra using extinction coefficients as calculated by the open-source BH-Mie code \citep{bohren2008absorption}.
\emph{YunMa} is fully integrated into the \emph{TauREx 3} retrieval platform \citep{Al-Refaie2022taurex,alrefaie2021taurex31} and, for the first time, provides cloud microphysics capabilities into a retrieval model. 
We describe the model in Section \ref{sec:model}. In Section \ref{sec:method}, we detail the experimental setups. In Section \ref{sec:result}, we validate particle size distributions and spectroscopic simulations against previous studies published in the literature. After validation, we show new spectral and retrieval simulations obtained with \emph{YunMa}. In Section \ref{sec:discussion}, we discuss our results and assumptions and identify possible model improvements to be considered in future developments.

\section{\emph{YunMa} description} \label{sec:model}
\emph{YunMa} estimates the vertical distribution of the cloud particle sizes (VDCP hereafter, see, e.g. Fig. \ref{fig:Gao_validation} b and \ref{fig:fsed_sensitivity} a, b) based on A-M model and their contribution to the radiative transfer calculations. The \emph{YunMa} module has been integrated into the \emph{TauREx 3} retrieval platform: the combined \emph{YunMa}-\emph{TauREx} model 
is able to constrain the VDCP from observed atmospheric spectra, as described in detail below. 

\subsection{Modelling the cloud particle size distribution} \label{subsection:formation}

\begin{figure}[ht!]
\plotone{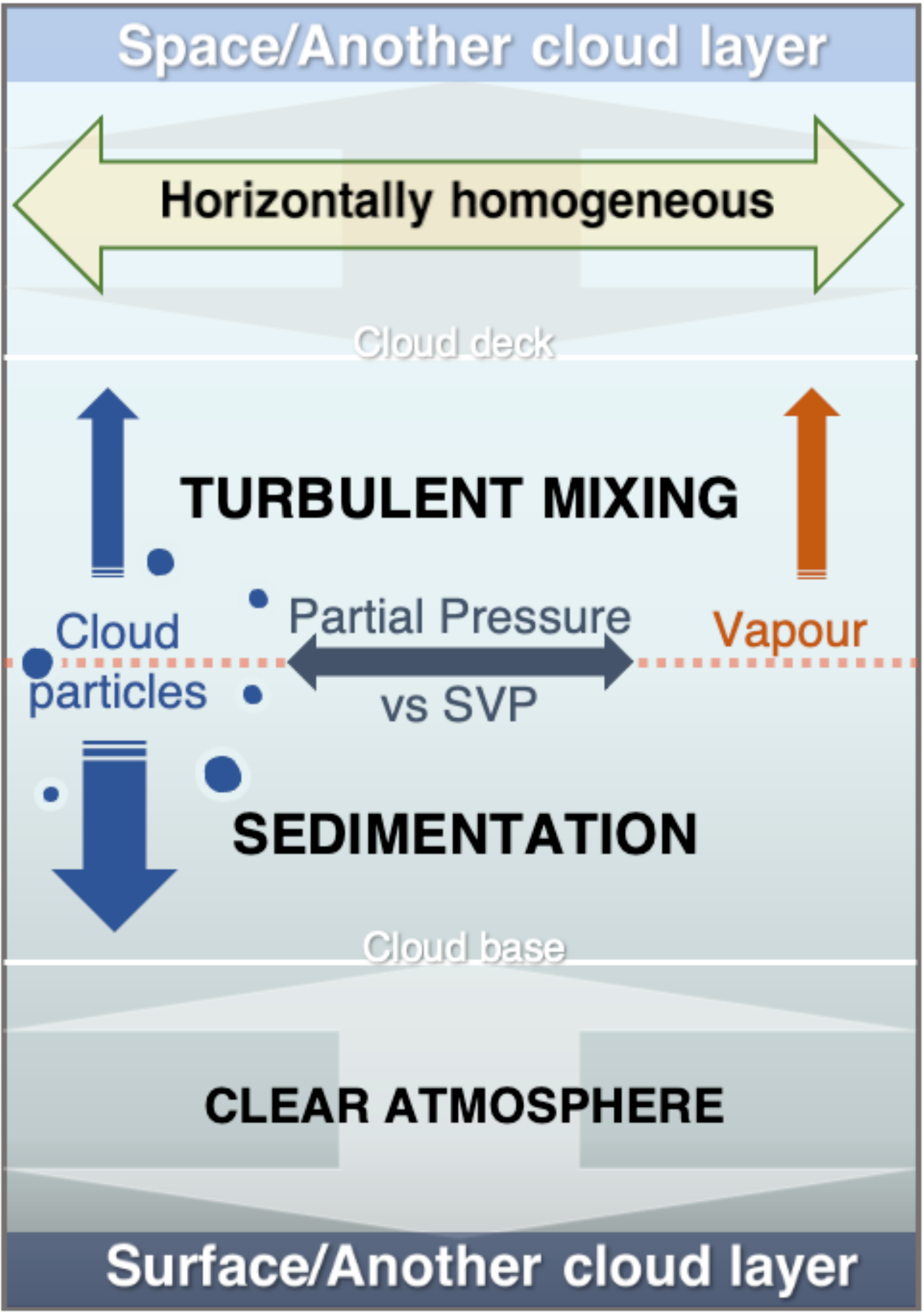}
\caption{Sketch of the A-M micophysical approach adopted in \emph{YunMa}. Cloud particles may form when the mixing ratio of the condensable gas exceeds the saturation mixing ratio, which is derived from its saturation vapour pressure (SVP). The vertical distribution of the cloud particle sizes (VDCP) is derived from the balance between the sedimentation of the cloud particles and the atmospheric turbulent mixing.
\label{fig:schematics}}
\end{figure}

\emph{YunMa} model contains a numerical realisation of the A-M microphysical approach to simulate the VDCP. We show in Fig. \ref{fig:schematics}  a pictorial representation of the A-M approach: it assumes that clouds form with different VDCP to maintain the balance between the upward turbulent mixing and downward sedimentation of the condensable species. Depending on the atmospheric $T$-$p$ profile, multiple cloud layers may form. 

\subsubsection{Cloud mixing profile} \label{subsection:qc}

Cloud particles start forming when the partial pressure of a certain gas exceeds the saturation vapour pressure (SVP):  the formation strongly depends on the atmospheric thermal structure. The condensation process, occurring when the partial pressure exceeds the SVP, is estimated by comparing the molecular mixing ratio of the gas phase with its saturation vapour mixing ratio: 

\begin{equation}
q_\mathrm{c}(z) = \max[0, \;q_\mathrm{t}(z) - (S_{\mathrm{c}} + 1) \;q_s(z)],
\label{eq:qc_condense}
\end{equation}
where $q_\mathrm{c}$ is the mixing ratio of the condensed species, $z$ is the altitude, $q_s$ is the mixing ratio where the condensable gas saturates, $q_\mathrm{t}$ is the total mixing ratio of a condensable chemical species, including both the condensate and gas phases, and $S_{\mathrm{c}}$ is the supersaturation factor which persists after condensation. $q_s$ can be estimated from the ratio between the SVP of a certain chemical species and the atmospheric pressure at the same altitude. Note that in this paper, the mixing ratio refers to the volume fraction of a chemical species in the atmosphere.

In the A-M approach, the turbulent mixing of the condensate and vapour is assumed to be in equilibrium with the sedimentation of the condensate:
\begin{equation}
-K(z)\;\frac{\mathrm{\partial} q_\mathrm{t}(z)}{\mathrm{\partial} z}-f_{\mathrm{sed}} \; w_\ast(z) \; q_\mathrm{c}(z)= 0,
\label{eq:main}
\end{equation}
where $K$ ($\mathrm m^2~\mathrm s^{-1}$) represents the vertical eddy diffusion coefficient, and $w_\ast$ ($\mathrm m~\mathrm s^{-1}$) is the convective velocity. $f_{\mathrm{sed}}$ is the ratio between the mass-weighted droplet sedimentation velocity and $w_\ast$, defined as:
\begin{equation}
f_{\mathrm{sed}}=\frac{\si{\int_0^\infty} v_f \frac{\mathrm{d}m}{\mathrm{d}r}\mathrm{d}r}{\varepsilon \rho_\mathrm{a} w_\ast q_\mathrm{c}};
\label{eq:fsed}
\end{equation}
here $\rho_\mathrm{a}$ is the atmospheric mass density, which can be estimated through the Ideal Gas Law, $\rho_\mathrm{p}$ is the mass density of a condensed particle, $\varepsilon$ is the ratio between the molecular weights of the condensates and the atmosphere, and $v_f$ is the sedimentation velocity which will be explained later. The first term in equation (\ref{eq:main}) describes the upward vertical draft derived from the macroscopic eddy diffusion equation. The second term describes the downward sedimentation, which is in equilibrium with the first term.

The eddy diffusion coefficient ($K$) is one of the key parameters affecting cloud formation. In free convection \citep{gierasch1985energy}, it can be estimated as:
\begin{equation}
K=\frac{H}{3} \left(\frac{L}{H}\right)^{\frac{4}{3}}\left(\frac{R F}{\mu \rho_\mathrm{a} c_\mathrm{p}}\right)^{\frac{1}{3}},
\label{eq:K}
\end{equation}
where $H$, $\mu$ and $c_\mathrm{p}$ are, respectively, the atmospheric scale height, mean molecular weight and specific heat capacity. $F = \sigma T_{\mathrm{eff}}^4$ is the approximated radiative flux. The turbulent mixing length ($L$) is the scale height of the local stability in eddy diffusion, as opposed to the atmospheric scale height ($H$). 
\emph{YunMa} has an application programming interface for $K$ and it can use the values provided by disequilibrium chemistry models plugged by the users into \emph{TauREx 3} -- e.g. the kinetic model plugin of \emph{TauREx 3} \citep{alrefaie2022freckll}, and derive $L$ accordingly, from equation (\ref{eq:K}). The convective velocity scale ($w_\star$) mentioned above can also be estimated as a ratio between $K$ and $L$. There are different ways to constrain $K$ from the atmospheric chemical and vertical advective time scales \citep[e.g.][]{Baeyens2021, Komacek_2019, parmentier2013, Zhang_2018}. However, the current estimations of $K$ in the literature lack validation from observations. While \emph{YunMa} is designed to be self-consistent with these approaches, this paper uses constant $K$ in the experiments as a first-order estimation. Free convection is justified by assuming the cloud forms in the deep convective layer of the atmosphere and by neglecting 3-D effects. These approximations in retrieval studies will need to be revisited with the improved quality of the data available and computing facilities.

The sedimentation velocity, denoted by $v_f$, is the speed at which a cloud particle settles within a heterogeneous mixture due to the force of gravity.
$v_f$ can be estimated through viscous fluid physics:
\begin{equation}
\label{eq:viscous}
v_f = \frac{2}{9} \frac{\beta g r^2 \Delta \rho}{\eta},
\end{equation}
where $\Delta \rho$ is the difference between $\rho_\mathrm{p}$ and $\rho_\mathrm{a}$, $\beta$ is the Cunningham slip factor, and  $\eta$ is the atmospheric dynamical viscosity (See Appendix \ref{appendix:eqA} for more details of SVP, $\beta$ and $\eta$).

\subsubsection{Particle size and number density} \label{subsection:r}

Following the A-M approach, we assume spherical cloud particles with radii $r$. The particle radius at $w_\star$, denoted as $r_w$, can be obtained using these relationships between $v_f$ and $w_\star$:
\begin{equation}
\label{eq:wstar}
v_f (r_w) = w_\star,
\end{equation}
and
\begin{equation}
\label{eq:rw}
v_f = w_\star (\frac{r}{r_w})^\alpha,
\end{equation}
where $\alpha$ corresponds to the sedimentation velocity decrease in viscous flows.
In A-M, the particle size distribution was constrained by in-situ measurements of Californian stratocumulus clouds, which followed a broad lognormal distribution. The assumption of lognormal distribution allows estimating the geometric mean radius ($r_\mathrm{g}$), the effective radius ($r_\mathrm{eff}$) and the total cloud particle number density ($N$), using the detailed definitions and derivations listed in Appendix \ref{appendix:eqB}.

\subsection{Cloud contribution in transit spectra} \label{subsection:radiative}
To estimate the wavelength-dependent cloud contribution to transit spectra, \emph{YunMa} 
adopts the scattering theory and absorption cross sections as described in \citet[][BH-Mie hereafter]{bohren2008absorption}, assuming spherical cloud particles. The cross-section of the cloud particles ($k_{\lambda}$) at each wavelength ($\lambda$) and particle size are estimated through the extinction coefficient ($Q_{\mathrm{ext}}$) of the corresponding wavelength and particle size, derived by BH-Mie from the refractive indices of the cloud particles:
\begin{equation}
\label{eq:xsec}
k_{\lambda} = Q_{\mathrm{ext}} \; \pi r ^ 2.
\end{equation}
We used the water ice refractive indices reported in \citet{warren2008refractive} for our simulations of the temperate super-Earth. We show some examples of atmospheres with water ice clouds in Section \ref{sec:result}. Post-experimental tests were conducted to avoid the contamination of liquid water particles. 

We simulate the cloud optical depths from the particle sizes and number densities along each optical path, which passes the terminator at altitude $z_\mathrm{ter}$, with a path length $s_{z_\mathrm{ter}}$ of each atmospheric layer: 
\begin{equation}
\label{eq:tau_cloud}
\tau_\lambda = \mathrm{\int}^{z_\mathrm{top}}_{z_\mathrm{ter}} \mathrm{\int}^{\infty}_{0} k_{\lambda} \; \frac{\mathrm{d}n}{\mathrm{d}{r}} \; \mathrm{d}{r} \; \frac{\mathrm{d}s_{z_\mathrm{ter}}}{\mathrm{d}z} \; \mathrm{d}z,
\end{equation}
where $n$ is the accumulated number density of particles with a radius smaller than $r$ and $z_{\mathrm{top}}$ is the altitude at the top of the atmosphere. The contribution of the clouds to the transit spectra, $\Delta F_{\mathrm{c}}$, can be estimated as:
\begin{equation}
\label{eq:contrib}
\Delta F_{\mathrm{c}} = \frac{2\mathrm{\int}_{z_{\mathrm{bottom}}}^{z_{\mathrm{top}}}(R_{\mathrm{p}}+z)(1-e^{-\tau_\lambda})\;\mathrm{d}z}{R_{\mathrm{s}}^2},
\end{equation}
where $z_{\mathrm{bottom}}$ is the altitude at $R_{\mathrm{p}}$. While \emph{YunMa} has the capability to include any customised cloud particle size distribution in the spectral simulations, in this paper, we aim at model testing and, for simplicity, we use a single radius bin, i.e. uniform cloud particle size $r_\mathrm{c}$ = $r_\mathrm{g}$ (see equation \ref{eq:rg} in Appendix \ref{appendix:eqB}) for each atmospheric layer in the radiative-transfer simulation.

\subsection{Cloud simulation validation}

\begin{figure}[ht!]
\plotone{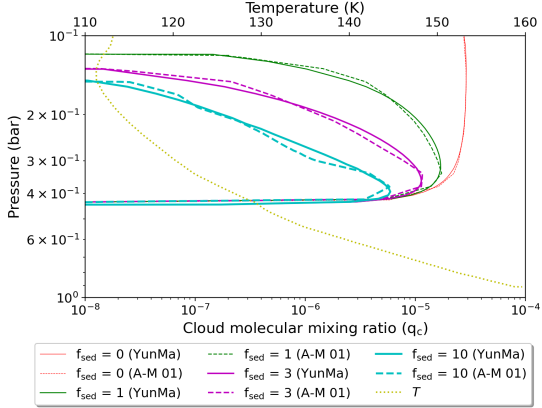}
\caption{Validation of the \emph{YunMa} cloud microphysics model against the Jovian ammonia clouds in \citet{ackerman2001precipitating}. $q_\mathrm{c}$ corresponding to different sedimentation efficiencies  ($f_{\mathrm{sed}}$) are shown. Solid lines: results from \emph{YunMa}. Dashed lines: results from Fig. 1 in \citet{ackerman2001precipitating}. Dotted line: $T$-$p$ profile.
\label{fig:AMvalidation}}
\end{figure}

We validated our implementation of A-M model against \cite{ackerman2001precipitating} by comparing the condensate mixing ratios of Jovian ammonia clouds ($q_\mathrm{c}$) with different values of $f_{\mathrm{sed}}$, as shown in Fig. \ref{fig:AMvalidation}. The two sets of results are consistent and the small differences in $q_\mathrm{c}$ translate into $\sim10^{-2}$ ppm in transit depth, which is completely negligible compared to typical observational noise. 

\begin{figure}[ht!]
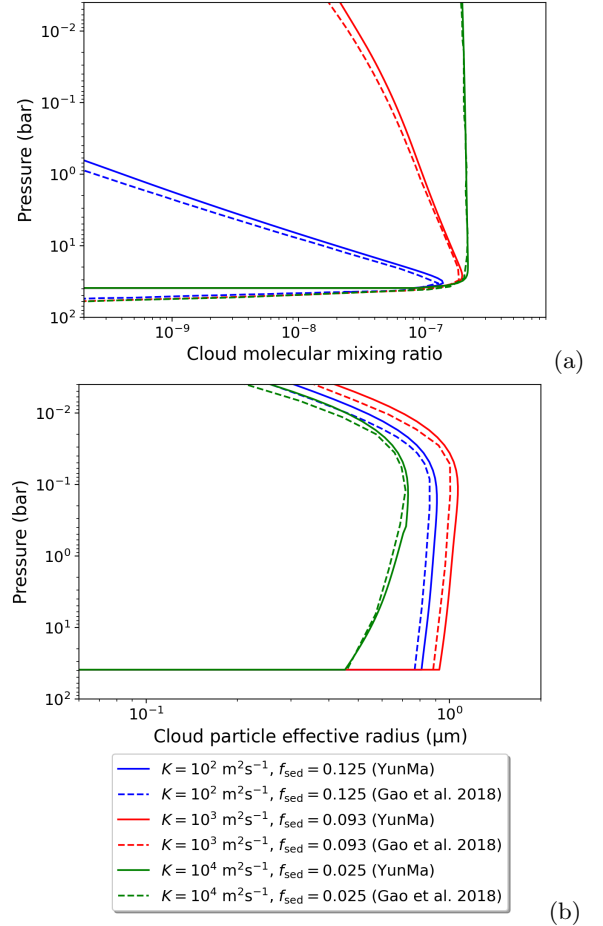

\begin{center}
\plotone{Gao_qc_Validation_updated.pdf}(a)
\plotone{Gao_reff_Validation_updated.pdf}(b)
\caption{Validation of \emph{YunMa} (solid lines) against the A-M model in \cite{gao2018sedimentation} (dashed lines). Top: condensate mixing ratios. Bottom: cloud particle effective mean radii. 
\label{fig:Gao_validation}}
\end{center}
\end{figure}

We further validated our implementation against the results from \cite{gao2018sedimentation} by comparing the KCl cloud molecular mixing ratio and particle-size profile (Fig. \ref{fig:Gao_validation}). \cite{gao2018sedimentation} used CARMA to simulate cloud microphysics in exoplanets and brown dwarfs with $T_{\mathrm{eff}} = 400$ K and $\mathrm{\log}$ $g$ = $3.25$, $4.25$ and $5.25$ (in cgs units), corresponding to planetary masses of $0.72$ \si{M_J}, $8.47$ \si{M_J}, and $44.54$ \si{M_J}. In the best fit between CARMA to A-M model, the $f_{\mathrm{sed}}$ = $0.125$, $0.093$ and $0.025$ for the cases with $K$ = $\mathrm{10^2}$, $\mathrm{10^3}$ and $\mathrm{10^4}$ \si{m^2\;s^{-1}}, respectively. The mixing length is derived from constant eddy diffusion, as described in equation (\ref{eq:K}). The results agree with each other within 8$\%$, i.e. $\sim10^{-3}$ ppm difference in the transit depth.

\begin{figure}[ht!]
\plotone{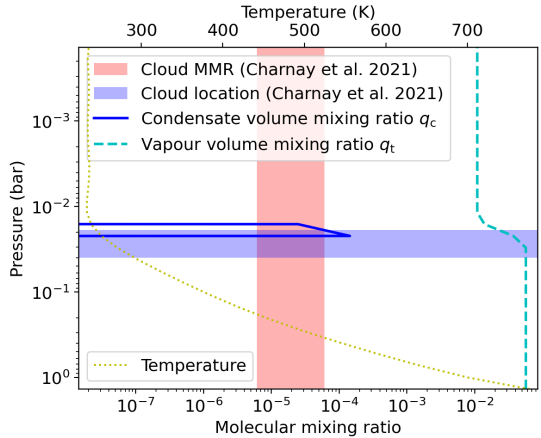}
\caption{Comparison of \emph{YunMa}'s results against simulations by \citet{charnay2021formation} for the 100 $\times$ solar metallicity scenario at the substellar point (see Fig. 6 a in the original paper). The condensate (blue) and vapour (cyan) MMRs are simulated by \emph{YunMa} using the A-M approach with $K=10^2$ m$^2$s$^{-1}$. The cross-section of the two shaded areas indicates the range of cloud MMR and location simulated by Laboratoire de Météorologie Dynamique Generic GCM (LMDG) in \citet{charnay2021formation}. LMDG is derived from the LMDZ Earth \citep{Hourdin2006LMDZ} and Mars \citep{Forget1999Mars} GCMs. \label{fig:Charnayvalidation}}
\end{figure}

We compared simulations from \emph{YunMa} with the results from \citet{charnay2021formation}, which include the horizontal effects generated by global circulation. While very precise comparison and validation are not possible in this case due to the two approaches' very different natures and assumptions, it is useful to test whether we can reproduce similar results when we use consistent assumptions. In the comparison shown in Fig. \ref{fig:Charnayvalidation}, we used the value of $K = 10^2$ \si{m^2s^{-1}} estimated in \citet{charnay2021formation} assuming 100 $\times$ solar composition. When setting $f_{\mathrm{sed}}$ to 3, \emph{YunMa} produced similar results to those reported by \citet{charnay2021formation}, with clouds forming in the region between 3 $\times$ 10$^{-2}$ and 1 $\times$ 10$^{-2}$ bar and a cloud molecular mixing ratio of approximately 10$^{-4}$.

\begin{figure*}[ht!]
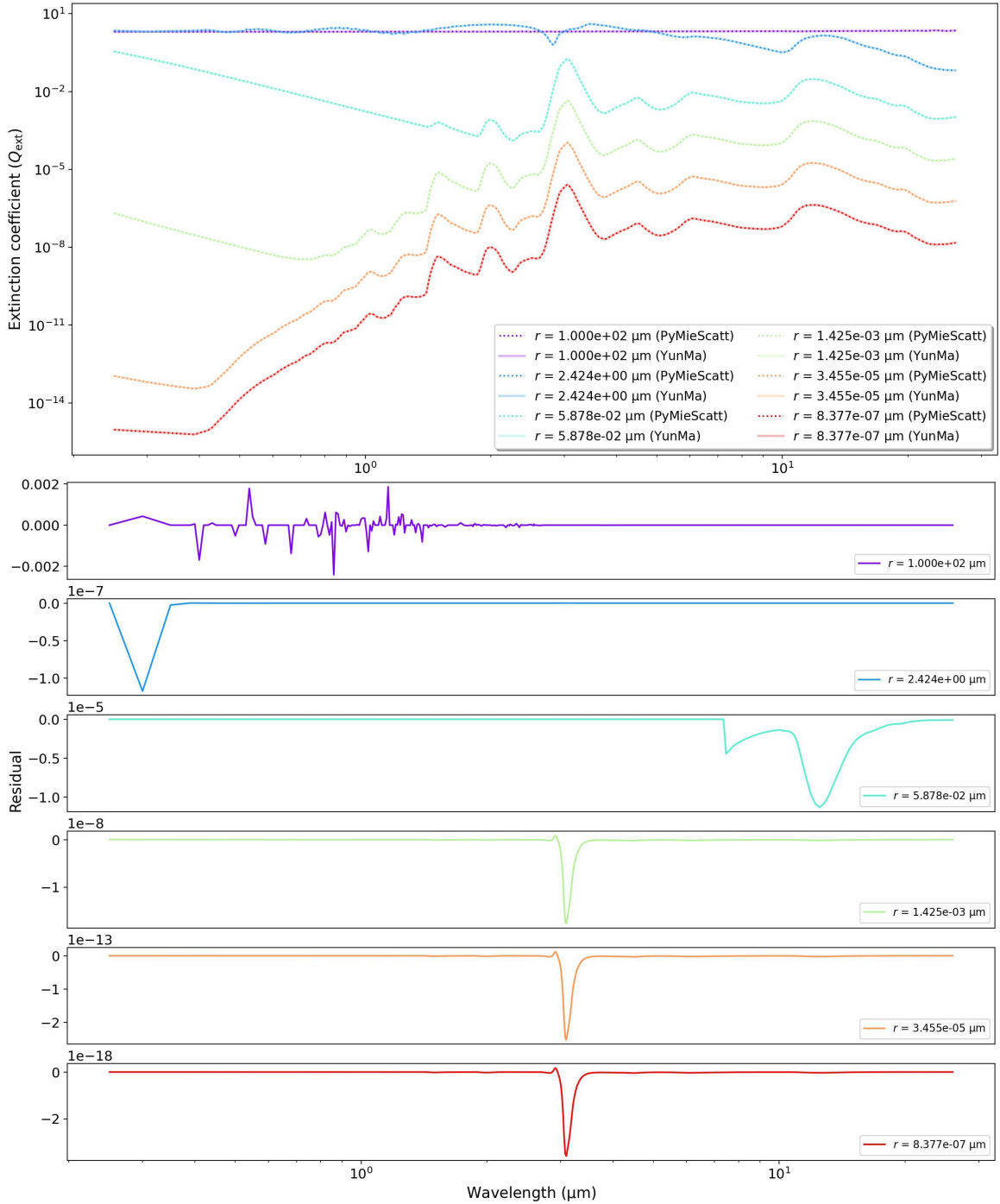

\plotone{pymiescatt_comp.pdf}
\plotone{residual_radii.pdf}
\caption{Validation of radiative transfer simulations obtained with \emph{YunMa} against the open source code PyMieScatt \citep{pymiescatt2018}. The extinction coefficients for cloud particles with different sizes are estimated from the theory of \cite{bohren2008absorption}. To address the computational limitations of retrievals, we pre-calculated the extinction coefficients used in equation (\ref{eq:xsec}) to estimate the cross-sections of the cloud particles. The pre-calculated list includes values for particle radii from 1 $\times$ 10$^{-7}$ to 1 $\times$ 10$^{-2}$ \si{\micro m}, equally spaced in the logarithm space. Here we show only six examples in the top panel. Bottom panels: residuals obtained by subtracting the extinction coefficients as estimated by the two codes, \emph{YunMa} and PyMieScatt. The maximum discrepancy, corresponding to the largest particle radius simulated here ($r$ = 1 $\times$ 10$^{2}$), is negligible, i.e. $\pm$0.002. \label{fig:residual}}
\end{figure*}

We have validated the BH-Mie module in \emph{YunMa} against PyMieScatt, an open-source model simulating atmospheric particle scattering properties \citep{pymiescatt2018}, as shown in Fig. \ref{fig:residual}. Cloud particle radii were selected in the range of 0.1-100 \si{\micro m}. The largest discrepancy in $Q_{\mathrm{ext}}$ is within $\pm$0.002, which corresponds to an average of 0.01 ppm in the planetary transit depth of the nominal scenario in our experiments.

\subsection{YunMa-TauREx: retrieval of cloudy atmospheres}
We integrate the \emph{YunMa} VDCP and $\tau_\lambda$ simulations in the Tau Retrieval of Exoplanets framework \citep[\emph{TauREx 3},][]{Al-Refaie2022taurex, alrefaie2021taurex31}, which allows atmospheric retrieval simulations. \emph{TauREx 3} combined to \emph{YunMa} allow us to perform retrievals which include cloud microphysical processes and cloud scattering properties. Parameters estimated by \emph{TauREx 3} include atmospheric $T$-$p$ and chemical profiles, planetary (e.g., mass and radius) and stellar (e.g., temperature and metallicity) parameters. The radiative transfer calculations executed by \emph{TauREx 3} consider molecular and atomic absorptions, Rayleigh scattering and collisionally induced absorptions (CIA) of H$_2$-H$_2$ and H$_2$-He pairs from \cite{Cox2015allen}. 

\emph{YunMa} uses as initial condition the gas mixing ratio profiles provided by \emph{TauREx 3} chemistry models ($q_\mathrm{t}$, equation \ref{eq:qc_condense}, \ref{eq:main}). In this paper, for simplicity, we assume the baseline chemical abundances are constant with altitude instead of a more complex chemical structure. \emph{YunMa} then adjusts the gas phase mixing ratios, atmospheric mean molecular weight and atmospheric density in the \emph{TauREx 3} chemistry models as a result of the formation of clouds. To simulate transit spectra and perform retrievals, we use the atmospheric grids and optical paths defined in \emph{TauREx 3} and add the cloud opacities as estimated by \emph{YunMa} BH-Mie to the absorptions caused by the chemical species, using the methods explained in Section \ref{subsection:radiative}. The retrievals were tested on 80 Intel(R) Xeon(R) Gold 6248 CPU @ 2.50GHz.


\section{Methodology} \label{sec:method}

\begin{deluxetable}{lllll}
\tablecaption{Priors for spectral retrieval experiments using \emph{YunMa} of all the cases listed in Table \ref{tab:retrieval}. \label{tab:priors}}
\tabletypesize{\scriptsize}
\tablehead{\colhead{Parameter} & \colhead{Unit} & \colhead{Ground} & \colhead{Mode} & \colhead{Priors}
}
\startdata
$R_{\mathrm{p}}$ & \nomJovianEqRadius & 0.20 & factor & 0.75 -- 1.25
\\ 
$f_{\mathrm{sed}}$ & \nodata & Table \ref{tab:retrieval} & log & 10$^{-3}$ -- 10$^2$
\\
$X_{\mathrm{H_2O}}$ & \nodata & Table \ref{tab:retrieval} & log & 10$^{-12}$ -- 1
\\
\hline
$p_{\mathrm{c}}$ & \si{bar} & Table \ref{tab:retrieval} & log & 10$^{-4}$ -- 1
\\
$T_{\mathrm{c}}$ & \si{K} & 200 & linear & 0 -- 500
\\
$T_{\mathrm{surf}}$ & \si{K} & 1000 & linear & 500 -- 2000
\\
\hline
$X_{\mathrm{N_2}}$ & \nodata & Table \ref{tab:retrieval} & linear & 10$^{-12}$ -- 1 \\
\enddata
\tablecomments{$X_{\mathrm{H_2O}}$ represents the water vapour mixing ratio}
\end{deluxetable}


In this paper, we use \emph{YunMa} to perform retrieval simulations of small temperate planets, where we expect a considerable amount of H$_2$O to be present in the atmosphere. For simplicity, we consider only water clouds forming in the atmosphere and we do not consider supersaturation cases. The planetary parameters are inspired by K2-18 b \citep{tsiaras2019water, charnay2021formation, yu2021identify}, which is a suitable candidate for cloud model testing. We list all the priors of our experiment in Table \ref{tab:priors}. In this work, we estimate $\eta$ using the approximation proposed by \citet[][equation \ref{eq:rosner}]{rosner2012transport}. 
We include both scattering and absorption due to water clouds based on 
\citet{bohren2008absorption}, Rayleigh scattering of all the gas species and CIA of H$_2$-H$_2$ and H$_2$-He pairs, which are enabled by \emph{TauREx 3}. We use N$_2$ as a representative inactive gas undetectable spectroscopically but that contributes to the increase of the atmospheric mean molecular weight, $m$, and decrease of scale heights $H = k_\mathrm{B}T/mg$. H$_2$ and He act as the filling gases. We use the POKAZATEL dataset for $^1$H$_2^{16}$O \citep{Polyansky2018pokazatel} from the ExoMol database\footnote{https://exomol.com} \citep{tennyson2012exomol,Tennyson2021exomol,chubb2022exomol} to estimate the water vapour absorption and Rayleigh scattering. The CIA data is from HITRAN\footnote{https://hitran.org} \citep{Karman2019hitran}. We use the PHOENIX library \citep{Husser2013phoenix} to simulate the stellar atmospheres spectra.

For the numerical parameter settings, after a number of tests, we decided to use the explicit Runge-Kutta method of order 8 \citep[DOP853,][]{Hairer1988} with relative tolerance ($rtol$) of 1 $\times$ 10$^{-13}$ and absolute tolerance ($atol$) of 1 $\times$ 10$^{-16}$ to solve the partial differential equation (\ref{eq:main}) for all the experiments presented in Section \ref{sec:result}.
We have opted for a logarithm sampling to retrieve most of the atmospheric parameters, e.g., $f_{\mathrm{sed}}$, $X_{\mathrm{H_2O}}$ and $p_{\mathrm{c}}$. We have used a linear sampling, instead, for N$_2$ to obtain a better numerical performance. The priors are sufficiently unconstrained to avoid biases generated by excessive pre-knowledge, as discussed, e.g. in \citet{changeat2021bias}. After a number of tests, we have chosen to use 400 live points for 3-dimensional retrievals and 1000 for more dimensions.


To begin with, we run a sensitivity study with \emph{YunMa} about the planetary and instrumental parameters. We set the planetary radius ($R_\mathrm{p}$), $f_{\mathrm{sed}}$ and $X_{\mathrm{H_2O}}$ as free parameters in our 3-dimensional retrieval tests. We list in Table \ref{tab:priors} the planetary parameters adopted in the simulations, the prior ranges and the sampling modes. The simulations are conducted with 80 atmospheric layers, from 10 \si{bar} to $10^{-6}$ \si{bar}, which encompass the typical observable atmospheric range for super-Earths. We select Case \ref{case:p200} in Table \ref{tab:retrieval} as the nominal case, and test the model sensitivity to the key parameters in the retrievals. In Case \ref{case:p200}, clouds dampen the gas spectroscopic features but do not obscure them entirely (see Fig. \ref{fig:forwardcloud}). The nominal $f_{\mathrm{sed}}$ refers to the value adopted in A-M.  The water SVP (both liquid and ice) used are taken from Appendix A in \citet{ackerman2001precipitating}. In this experiment, we will perform sets of retrievals with the aim of:
\begin{enumerate}
    \item Sensitivity studies to key atmospheric parameters (Case \ref{case:p100}--\ref{case:p200H2O05})
    \item Sensitivity studies to data quality (Case \ref{case:p200}, \ref{case:p200err1}--\ref{case:p200wl1})
    \item Retrievals of atmospheric thermal profiles (Case \ref{case:p200Tp}--\ref{case:p1000Tp})
    \item Addition of N$_2$ in retrievals (Case \ref{case:p200N2m12}--\ref{case:p200N205})
    \item Degeneracy between clouds and heavy atmosphere (Case \ref{case:p200N205fix05}--\ref{case:p200N205fix0})
    \item Comparison of cloud retrieval models (Case \ref{case:p200N205Tp}, \ref{case:p200N205simple}--\ref{case:p200N205nocloud})
    \item Retrievals of featureless spectra (Case \ref{case:p500N201}--\ref{case:p500N2051ppm})
\end{enumerate}

\subsection*{T-p profile}

Isothermal $T$-$p$ profiles, as commonly used in transit retrieval studies, are too simplistic for cloud studies. In our experiments, we first assume $T$-$p$ profiles with a dry adiabatic lapse rate in the troposphere, a moist adiabatic lapse rate in the cloud-forming region and a
colder isothermal profile above the tropopause. However, to be compatible with the computing requirements in retrieval, we simplify it to a \emph{TauREx} ``two-point'' profile \citep[e.g., Fig. \ref{fig:Tpfit} in Section \ref{sec:result}, ``N-point'' performances evaluated in][]{changeat2021bias}. We define as $T_{\mathrm{c}}$ and $p_{\mathrm{c}}$ the temperature and pressure at the tropopause and $T_{\mathrm{surf}}$ the temperature at 10 bar. 
The two-point profile is a fit to the points ($T_{\mathrm{c}}$, $p_{\mathrm{c}}$) and ($T_{\mathrm{surf}}$, $p_{\mathrm{surf}}$). One condition of cloud formation is that the atmospheric pressure exceeds the SVP, which is influenced by the thermal gradient in the lower atmosphere controlled by $T_{\mathrm{surf}}$ and $T_{\mathrm{c}}$. These factors determine the location where the pressure exceeds the SVP and therefore the cloud formation.

\subsection*{Instrumental performance}
The new generation of space-based facilities, such as JWST and Ariel, will deliver unprecedentedly high-quality data in terms of wavelength coverage, signal-to-noise ratio and spectral resolution. We select as nominal case transit spectra covering 0.4--14 \si{\micro m}, at a spectral resolution of 100, with 10 ppm uncertainty across wavelengths. We chose 0.4 \si{\micro m} as the blue cut-off to maximise the information content about Rayleigh scattering and 14 \si{\micro m} as the red cut-off to maximise the information content about water vapour and atmospheric temperature for the type of planets considered here. The choice of wavelength coverage and precision are inspired by current and planned instrumentation, while not trying to reproduce a specific observatory with its own limitations. The focus of this paper is on the retrievability of clouds and not on the performance of a specific facility.


\section{Results} \label{sec:result}
\subsection{Simulated transit spectra with YunMa} \label{subsection:forward}

\begin{figure*}[ht!]
\plotone{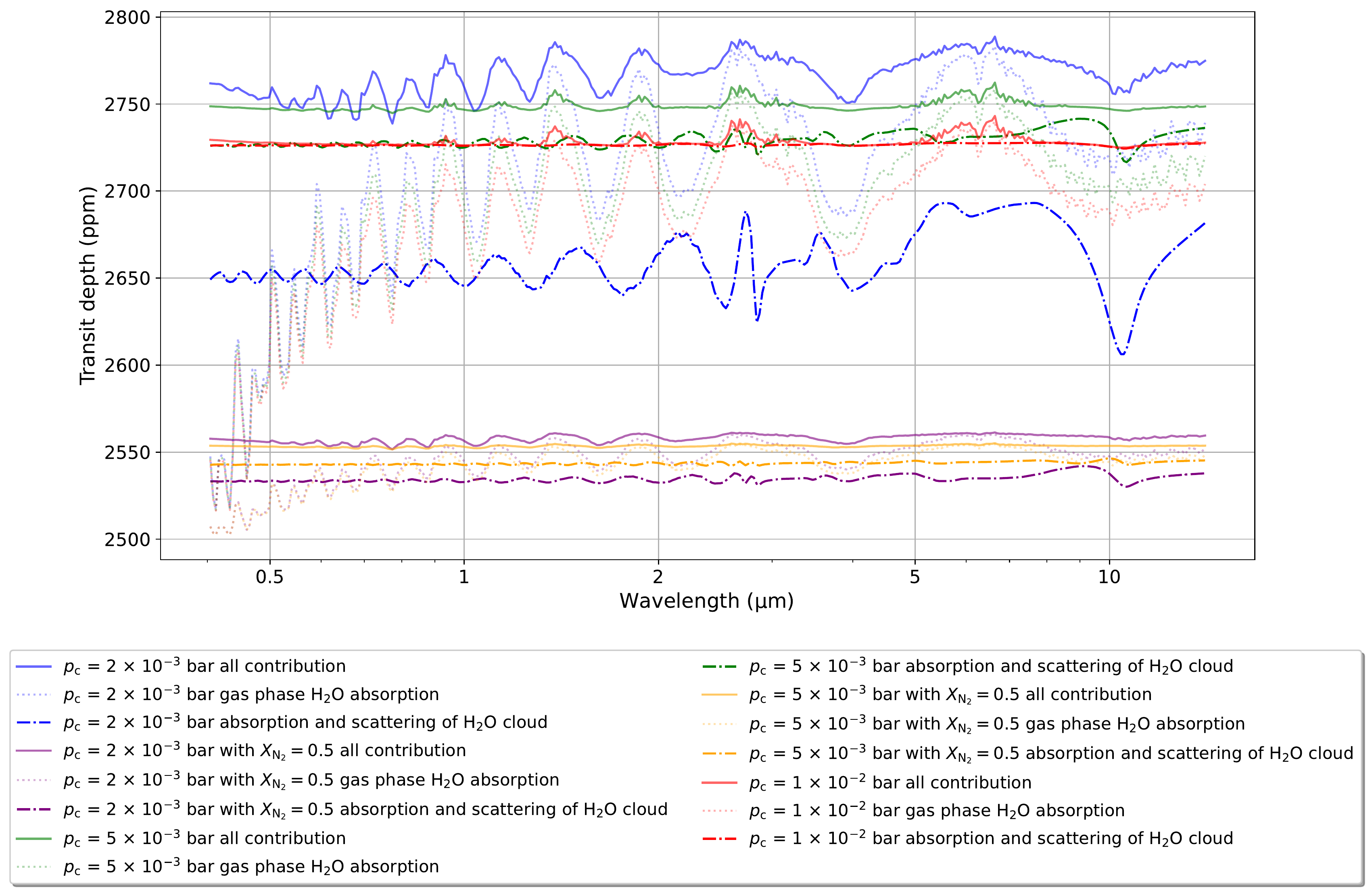}
\caption{Simulated transit spectra of cloudy super-Earths using \emph{YunMa}. Solid line: total transit depth with all contributions included; dot-dashed line: water ice clouds; faint-dotted line: water vapour. Blue, green and red lines: H$_2$/He dominated, cloudy atmospheres with different $p_{\mathrm{c}}$ (see legend). Purple and yellow lines: heavier cloudy atmospheres with 50$\%$ N$_2$.
\label{fig:forwardcloud}}.
\end{figure*}

We present here the transmission spectra generated with \emph{YunMa} of a cloudy super-Earth. Fig. \ref{fig:forwardcloud} shows five examples at resolving power of 100 which corresponds to the ground truths of some of the retrieval cases in Table \ref{tab:retrieval}: $p_{\mathrm{c}}$ = $2 \times 10^{-3}$ bar with $X_{\mathrm{N_2}}$ = 0 (blue, Case \ref{case:p200}, \ref{case:p200err1}, \ref{case:p200err30}, \ref{case:p200Tp} and \ref{case:p200Tperr1}) and = 0.5 (purple, Case \ref{case:p200N205}--\ref{case:p200N205nocloud}); $p_{\mathrm{c}}$ = $1 \times 10^{-2}$ bar with $X_{\mathrm{N_2}}$ = 0 (green, Case \ref{case:p500} and \ref{case:p500Tp}) and = 0.5 (yellow, Case \ref{case:p500N205} and \ref{case:p500N2051ppm}) and one case of opaque cloud with $p_{\mathrm{c}}$ = $1 \times 10^{-2}$ bar in the H$_2$/He dominated atmosphere (red, Case \ref{case:p1000} and \ref{case:p1000Tp}). The planetary and atmospheric parameters are listed in Table \ref{tab:priors} and \ref{tab:retrieval}. All the simulations contain baseline 10$\%$ H$_2$O abundance across the atmosphere, which is then altered by the cloud formation. The rest of the atmosphere is N$_2$ and H$_2$/He. We select $f_{\mathrm{sed}}$ = 3 for all the scenarios. The simulation results are summarised in Table \ref{tab:forward}.

\begin{deluxetable*}{llrrrrr}
\tablecaption{Atmospheric and cloud parameters included in \emph{YunMa} simulations. The corresponding transit depths are also reported. \label{tab:forward}}
\tabletypesize{\scriptsize}
\tablehead{\colhead{} & \colhead{$p_{\mathrm{c}}$} & \colhead{2 $\times$ 10$^{-3}$ (bar)} & \colhead{5 $\times$ 10$^{-3}$ (bar)} & \colhead{1 $\times$ 10$^{-2}$ (bar)} & \colhead{2 $\times$ 10$^{-3}$ (bar)} & \colhead{5 $\times$ 10$^{-3}$ (bar)}
\\
\colhead{} & \colhead{$X_{\mathrm{N_2}}$} & \colhead{\nodata} & \colhead{\nodata} & \colhead{\nodata} & \colhead{0.5} & \colhead{0.5}
\\
\hline
\colhead{Parameter} & \colhead{Unit} & \colhead{} & \colhead{} & \colhead{} & \colhead{} & \colhead{}
}
\startdata
All contribution, mean & ppm & 2770.17 & 2750 & 2729.82 & 2558.65 & 2553.71
\\ 
All contribution, std ($\sigma_{\mathrm{spec}}$) & ppm & 15.72 & 11.59 & 12.16 & 2.58 & 0.62
\\
Cloud contribution, mean & ppm & 2659.65 & 2729.60 & 2726.53 & 2534.78 & 2542.78
\\
Cloud contribution, std & ppm & 17.21 & 4.39 & 0.62 & 2.15 & 0.85
\\
MMW (bottom of the atmosphere) & g \si{mol^{-1}} & 3.88 & 3.88 & 3.88 & 16.73 & 16.73
\\
Atmospheric pressure (cloud base) & bar & 2.34 $\times$ 10 $^{-3}$ & 6.40 $\times$ 10 $^{-3}$ & 1.43 $\times$ 10 $^{-2}$ & 2.34 $\times$ 10 $^{-3}$ & 6.40 $\times$ 10 $^{-3}$
\\
Atmospheric pressure (cloud deck) & bar & 8.54 $\times$ 10 $^{-4}$ & 2.34 $\times$ 10 $^{-3}$ & 4.28 $\times$ 10 $^{-3}$ & 8.54 $\times$ 10 $^{-4}$ & 2.33 $\times$ 10 $^{-3}$
\\
Cloud MMR (cloud base) & \nodata & 6.35 $\times$ 10 $^{-4}$ & 9.51 $\times$ 10 $^{-5}$ & 2.00 $\times$ 10 $^{-4}$ & 2.92 $\times$ 10 $^{-4}$  & 4.35 $\times$ 10 $^{-4}$
\\
Cloud MMR (cloud deck) & \nodata & 7.56 $\times$ 10 $^{-8}$ & 2.92 $\times$ 10 $^{-7}$ & 7.28 $\times$ 10 $^{-8}$ & 3.49 $\times$ 10 $^{-7}$  & 1.33 $\times$ 10 $^{-6}$
\\
$v_f$ & m \si{s^{-1}} & 6.54 -- 7.53 & 6.07 - 6.67 & 4.43 -- 6.41 & 7.14 -- 7.97 & 4.37 -- 7.28
\\
$r_\mathrm{c}$ & \si{\micro m} & 9.03 -- 6.39 & 9.79 -- 8.88 & 12.13 -- 9.35 & 13.99 -- 11.09 & 25.53 -- 13.66
\\
$N$ (cloud base) & \si{m^{-3}} & 1.03 $\times$ 10$^4$ & 3.19 $\times$ 10$^4$ & 7.47 $\times$ 10$^4$ & 1.28 $\times$ 10$^4$ & 8.22 $\times$ 10 $^{3}$
\\
$N$ (cloud deck) & \si{m^{-3}} & 1.45 $\times$ 10$^1$ & 5.72 $\times$ 10$^1$ & 2.24 $\times$ 10$^1$ & 1.28 $\times$ 10$^1$ & 7.16 $\times$ 10 $^{1}$
\\
\enddata
\tablecomments{MMR and MMW:  molecular mixing ratio and  mean molecular weight. The pressure at the bottom of the atmosphere  is assumed to be 10 bar.}
\end{deluxetable*}

In the experiment without N$_2$ and $p_{\mathrm{c}}$ set to 2 $\times$ 10$^{-3}$ bar, clouds form at high altitude, where the atmospheric density ($\rho_\mathrm{a}$) is low compared to the cases where $p_{\mathrm{c}}$ = 5 $\times$ 10$^{-3}$ bar and = 1 $\times$ 10$^{-2}$ bar. Here the sedimentation velocity ($v_f$) is small with small cloud particle radii and number density. The cloud contribution (blue dash-dotted line) has a mean transit depth of 2660 ppm and $\sigma_{\mathrm{spec}}$ of 17 ppm. It is an optically thin cloud which does not completely block the spectral features shaped by water vapour absorption (blue dotted line). 
When $p_{\mathrm{c}}$ = 1 $\times$ 10$^{-2}$ bar,  clouds form at relatively low altitudes, where the atmospheric density ($\rho_\mathrm{a}$) is high. Here $v_f$ is large and the cloud particles have relatively large radii and number density, which increase the opacity. The cloud contribution (red dash-dotted line) has a mean flux depth of 2727 ppm and $\sigma_{\mathrm{spec}}$ of 0.62 ppm. Since the clouds are optically thick, they contribute significantly to the mean transit depth and obscure the spectral features of water vapour (red dotted line). However, the water vapour features are still able to show due to the low altitude of the clouds. Still, the spectral deviation is only 12.16 ppm, where the spectroscopic features have a high chance of being hidden by the observational uncertainty. 
$p_{\mathrm{c}}$ = 5 $\times$ 10$^{-3}$ bar is an intermediate case regarding the simulated cloud altitude and opacity. The simulation suggests that the intermediate combination of these two cloud properties does not result in more significant atmospheric features than in other cases.

In atmospheres with relatively high mean molecular weight -- and therefore small scale height -- for the same value of $p_{\mathrm{c}}$, the transit depth is smaller, as expected. In the case with $X_{\mathrm{N_2}}$ = 0.5 and $p_{\mathrm{c}}$ = 2 $\times$ 10$^{-2}$ and = 5 $\times$ 10$^{-2}$ bar, the mean value of the transit depths are $\sim200$ ppm smaller than them in a H$_2$/He dominated atmosphere. Here the cloud particles form at higher $\rho_\mathrm{a}$ and therefore have larger particle size and larger number density compared to those formed in the H$_2$/He dominated atmospheres. The spectrum has $\sigma_{\mathrm{spec}}$ of 2.58 and 0.62 ppm, which are negligible compared to the observational uncertainty.

Besides $p_{\mathrm{c}}$ and $X_{\mathrm{N_2}}$, we also have tested different $f_{\mathrm{sed}}$ to understand how this parameter controls the cloud microphysics. The particle radii, $r_{\text{c}}$, number density and transit spectra across all the cloud pressure levels and obtained with different $f_{\mathrm{sed}}$ are shown in Fig. \ref{fig:fsed_sensitivity}. Here we note that, from the results, the cloud particle sizes increase with $f_{\mathrm{sed}}$ while the number densities at each layer behave reversely. Also, which is easy to understand, the more atmospheric layers with clouds, the larger the optical depth.

\subsection{Retrieval results}
We show in this section how \emph{YunMa}  performs with different model assumptions and ground truth parameters (GTPs), following the approach described in Section \ref{sec:method}. GTPs and priors are listed in Table \ref{tab:priors}. The retrieved values and one standard deviation (1$\sigma$) of the posterior distributions obtained for all the simulated cases are summarised in Table \ref{tab:retrieval}.

\subsubsection*{Sensitivity studies to key atmospheric parameters}
\begin{figure}[ht!]
\plotone{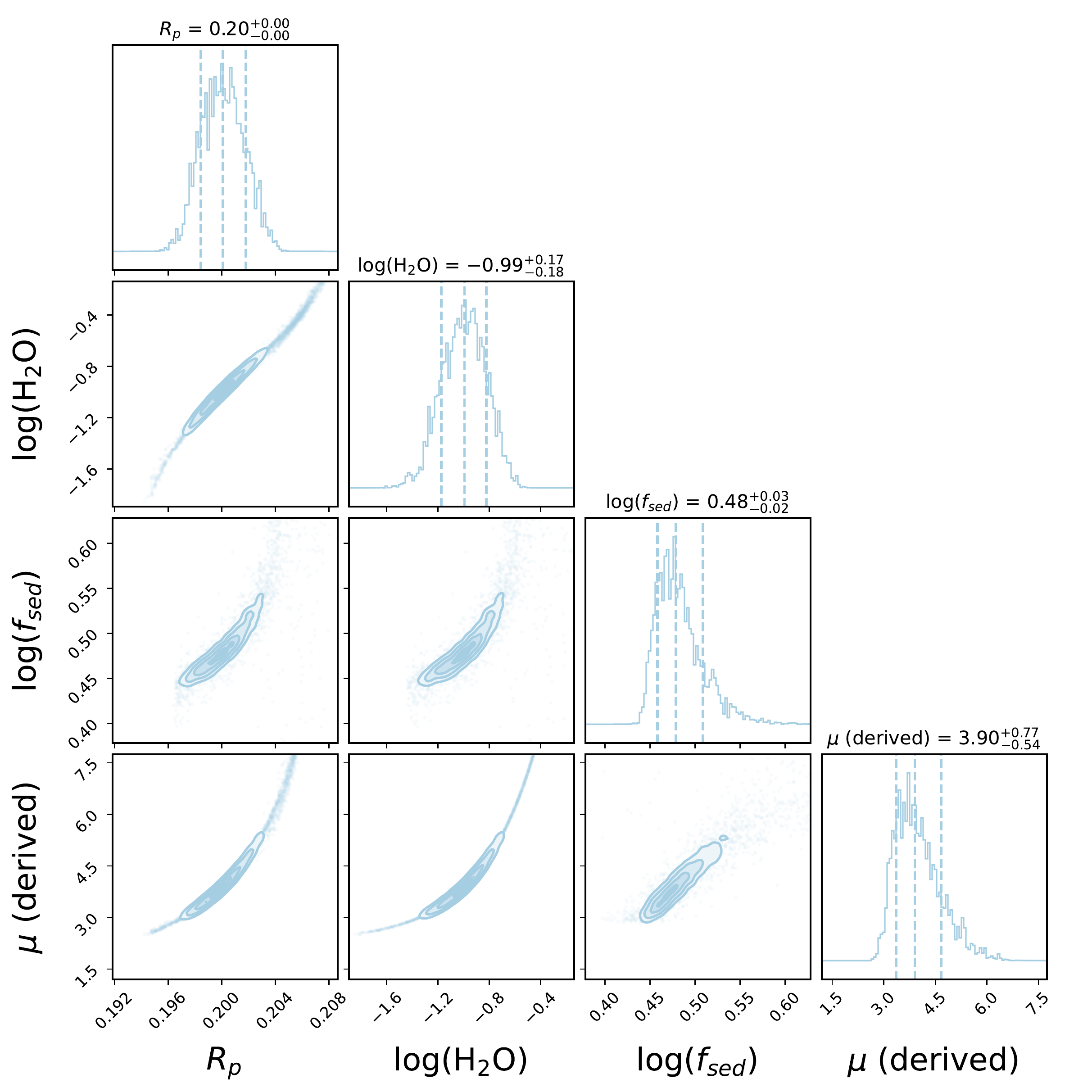}
\caption{Example of cloud retrieval using \emph{YunMa} integrated in \emph{TauREx 3} (Case \ref{case:p200} in Table \ref{tab:retrieval}). The sedimentation efficiency ($f_{\mathrm{sed}}$), which is the main parameter controlling the cloud microphysics, is well recovered together with the vapour water mixing ratio.
\label{fig:2Dposterior}}
\end{figure}

We have performed sensitivity studies to test how the model behaves when changing the key atmospheric parameters, including $p_{\mathrm{c}}$, $f_{\mathrm{sed}}$, $X_{\mathrm{H_2O}}$, $X_{\mathrm{N_2}}$ (Case \ref{case:p100}--\ref{case:p200H2O05}). 
Cases \ref{case:p100}--\ref{case:p10000} test the effects of different $p_{\mathrm{c}}$ to the transit spectra and retrievals. Tuning $p_{\mathrm{c}}$ alters the cloud altitude and the optical thickness: in these cases, the larger is $p_{\mathrm{c}}$, the more opaque meanwhile, the lower altitude becomes the clouds. The significance of spectroscopic features owns to both factors. Generally speaking, the more significant features are, the easier it is to retrieve the atmospheric parameters. Here we test $p_{\mathrm{c}}$ from 10$^{-3}$ to 10$^{-1}$ bar, which is a much broader range than the one considered in previous literature about K2-18 b. The transit spectra of Cases \ref{case:p200} and \ref{case:p1000} are shown in Fig. \ref{fig:forwardcloud} (blue and red solid lines, respectively). We choose $p_{\mathrm{c}}$ = 2 $\times$ 10$^{-3}$ bar as nominal case and show the corresponding posterior distributions in Fig. \ref{fig:2Dposterior}. 

Cases \ref{case:p200}, \ref{case:p200fsed001} and \ref{case:p200fsed10} test the impact on the transit spectra and retrievals of the sedimentation efficiency ($f_{\mathrm{sed}}$), which controls the cloud microphysics in the model. In Case \ref{case:p200fsed001}, we set the sedimentation efficiency $f_{\mathrm{sed}}$ to 0.01, i.e. the downward sedimentation of the cloud particles is relatively slow compared to the net upward molecular mixing of the condensable species. By contrast, in Case \ref{case:p200fsed10} where $f_{\mathrm{sed}}$ = 10, we have a larger downward draft velocity scale compared to the upward one. 
In analysing the results, we utilize the term ``accuracy'' to indicate that our retrieved result is in a certain range of the ground truth and ``precision'' to the 1$\sigma$ of the posteriors. In the simple cases, when $f_{\mathrm{sed}}$ = 0.01 and 3 (Case \ref{case:p200} and \ref{case:p200fsed001}), the accuracy levels of $f_{\mathrm{sed}}$, $X_{\mathrm{H_2O}}$ and $R_{\mathrm{p}}$ are $>$ 90$\%$. In comparison, in the scenario where $f_{\mathrm{sed}}$ = 10 (Case \ref{case:p200fsed10}), the accuracy level is $>$ 70$\%$.
This result is not unexpected, as high $f_{\mathrm{sed}}$ scenarios tend to have negligible impact on the planetary transit depth due to thinner cloud layers and smaller number density ($N$) compared to a low $f_{\mathrm{sed}}$ scenario, e.g., the $f_{\mathrm{sed}}$ = 10 and 100 cases in Fig. \ref{fig:fsed_sensitivity}.

In Case \ref{case:p200H2O001} and \ref{case:p200H2O05}, we modulate the amount of condensable gas, here represented by the water vapour mixing ratio ($X_{\mathrm{H_2O}}$). In the cases of our experiment, the clouds start to form when $X_{\mathrm{H_2O}}$ reaches the significance of 1 $\times$ 10$^{-3}$. When $X_{\mathrm{H_2O}}$ = 0.01 (Case \ref{case:p200H2O001}), a thin cloud with low opacity may form, the water vapour spectral features are visible and the retrieved values of log($X_{\mathrm{H_2O}}$) and log($f_{\mathrm{sed}}$) have $>$ 90$\%$ accuracy. By contrast, a higher mixing ratio of the condensable gas increases the partial pressure and contributes to the condensation process. This is, for instance, the case of $X_{\mathrm{H_2O}}$ = 0.5 (Case \ref{case:p200H2O05}) where the cloud is thick and largely blocks the spectral features, making the retrieval of the atmospheric parameters difficult.

\subsubsection*{Sensitivity studies to data quality}
We then test how \emph{YunMa}'s performances degrade when we compromise with the data quality, for instance:
uncertainties of 30 ppm in Case \ref{case:p200err30} and spectral resolution of 10 in Case \ref{case:p200res10}, which should have similar effects.
We move the blue cut-off at longer wavelengths in Case \ref{case:p200wl1}.
From the experiments on observational data quality, Case \ref{case:p200err1}'s Bayesian evidence (4568.52) compared to the ones calculated for Case \ref{case:p200} (3755.80) and Case \ref{case:p200err30} (3368.29) showcases how the retrieval performance improves when the observational uncertainties are small. Case \ref{case:p200}'s performance surpasses Case \ref{case:p200res10}, as the spectral resolution of the transit spectrum used as input to the retrieval is higher. In Case \ref{case:p200wl1}, we omitted the information in the optical wavelengths, which means we have less information about the cloud scattering properties. The retrieval performances are degraded compared to Case \ref{case:p200}, which includes the optical wavelengths.

\subsubsection*{Retrievals of atmospheric thermal profiles}

In Case \ref{case:p200Tp}--\ref{case:p1000Tp}, we retrieved the $T$-$p$ profiles as free parameters for different $p_{\mathrm{c}}$ to test how \emph{YunMa} performs with increasingly complex model assumptions and which parameters may be problematic in these retrievals. Our results show that both the $T$-$p$ profiles and cloud parameters can be constrained, although the retrieved gas phase mixing ratio and $f_{\mathrm{sed}}$ distribution may have large standard deviations in some of the cases.

\subsubsection*{Addition of N$_2$ in retrievals}
Inactive, featureless gases, such as N$_2$, inject much uncertainty in the retrieval. In Cases \ref{case:p200N2m12}--\ref{case:p200N205}, we include different amounts of the inert and featureless gas N$_2$ in the atmosphere. N$_2$ may exist in super-Earths' atmospheres, as it happens for Solar System planets at similar temperatures. Being N$_2$ heavier than H$_2$O, we adjust the mean molecular weight and the scale heights accordingly by modulating $X_{\mathrm{N_2}}$.
Heavier atmospheres have smaller scale heights: the spectral features are less prominent and harder to detect. We first try to retrieve only $R_{\mathrm{p}}$, $f_{\mathrm{sed}}$, $X_{\mathrm{H_2O}}$ and $X_{\mathrm{N_2}}$.
The results show that despite the minimal spectral features of the cloudy heavy atmospheres due to N$_2$ injection, \emph{YunMa} is still able to retrieve these atmospheric parameters in simple cases.

\subsubsection*{Degeneracy between clouds and heavy atmosphere}

\begin{figure}[ht!]
\plotone{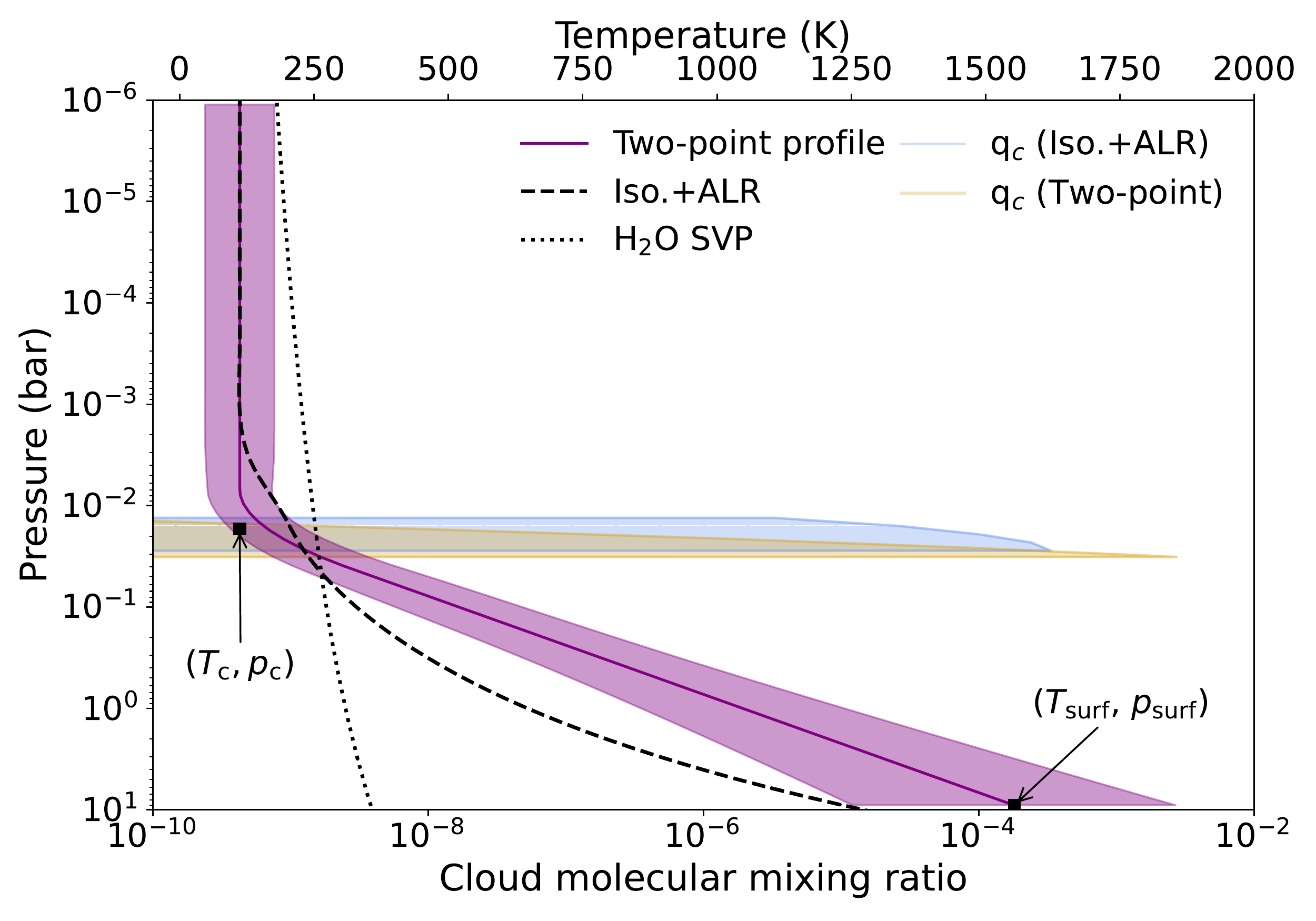}
\caption{Retrieved $T$-$p$ profile of Case \ref{case:p200N205Tp} in Table \ref{tab:retrieval}. The isothermal temperature ($T_{\mathrm{c}}$), surface temperature ($T_{\mathrm{surf}}$) and the pressure where the isothermal profile starts ($p_{\mathrm{c}}$) are retrieved using \emph{YunMa} and indicated by squares in the plot. The solid purple line indicates the two-point profile fitted from the retrieved $T_{\mathrm{c}}$, $T_{\mathrm{surf}}$ and $p_{\mathrm{c}}$ values. The shaded area indicates the standard deviation of the posterior distribution. We also plot a $T$-$p$ profile with an isothermal upper atmosphere, DALR in the lower atmosphere and MALR around the cloud-forming region to show how the two-point profile can deviate from the ALR when similar cloud cover forms. The two-point profile is a useful approximation to estimate cloud formation while reducing the retrieval computing time.\label{fig:Tpfit}}
\end{figure}

In Cases \ref{case:p200N205fix05}--\ref{case:p200N205nocloud}, we conduct more complex retrievals to test the degeneracy between clouds and $X_{\mathrm{N_2}}$. The corresponding transit spectra are shown in Fig. \ref{fig:forwardcloud} (purple lines). Similar to N$_2$, the existence of clouds mitigates the spectral features, and the difference between these two scenarios may be difficult to distinguish from the current data quality. Case \ref{case:p200N205fix05} retrieves these parameters except $X_{\mathrm{N_2}}$ fixed to 50$\%$ for comparison with Case \ref{case:p200N205Tp} to investigate the degeneracy imposed by the uncertainty of $X_{\mathrm{N_2}}$. The result of Case \ref{case:p200N205fix05} shows how the atmospheric parameters, with the exception of $X_{\mathrm{N_2}}$, can be retrieved in a heavy atmosphere with $X_{\mathrm{N_2}}$ fixed to the ground truth. When we include the uncertainty of $X_{\mathrm{N_2}}$ (Case \ref{case:p200N205Tp}), the GTPs for $R_{\mathrm{p}}$, $X_{\mathrm{H_2O}}$, $f_{\mathrm{sed}}$, $T_{\mathrm{surf}}$ and $T_{\mathrm{top}}$ still fall into the 2$\sigma$ confidence range, where $R_{\mathrm{p}}$, $X_{\mathrm{H_2O}}$ and $f_{\mathrm{sed}}$ have accuracy levels $>$ 60$\%$ and $T_{\mathrm{surf}}$ and $T_{\mathrm{top}}$ $>$ 45$\%$. An illustration of the retrieved $T$-$p$ profile is shown in Fig. \ref{fig:Tpfit}. $f_{\mathrm{sed}}$ is significantly less constrained in Case \ref{case:p200N205Tp} than Case \ref{case:p200N205fix05} due to the uncertainty of $X_{\mathrm{N_2}}$.

One hypothesis is that if we do not include N$_2$ among the priors, the model will add clouds to compensate for the missing N$_2$. In Case \ref{case:p200N205fix0}, we test this hypothesis by forcing $X_{\mathrm{N_2}}$ to zero and then monitor the cloud parameters in the posteriors obtained. The results suggest that potential degeneracy could happen, plotted in Fig. \ref{fig:N2simple}: when we omit N$_2$ among the priors, the retrieval tries to compensate for the missing radiative-inactive gas by decreasing $R_{\mathrm{p}}$ and $T_{\mathrm{surf}}$, while increasing $X_{\mathrm{H_2O}}$, and the Bayesian evidence of Case \ref{case:p200N205fix0} (3757.08) is close to Case \ref{case:p200N205Tp} (3757.40). 

\subsubsection*{Comparison of cloud retrieval models}
\begin{figure*}[ht!]
\plotone{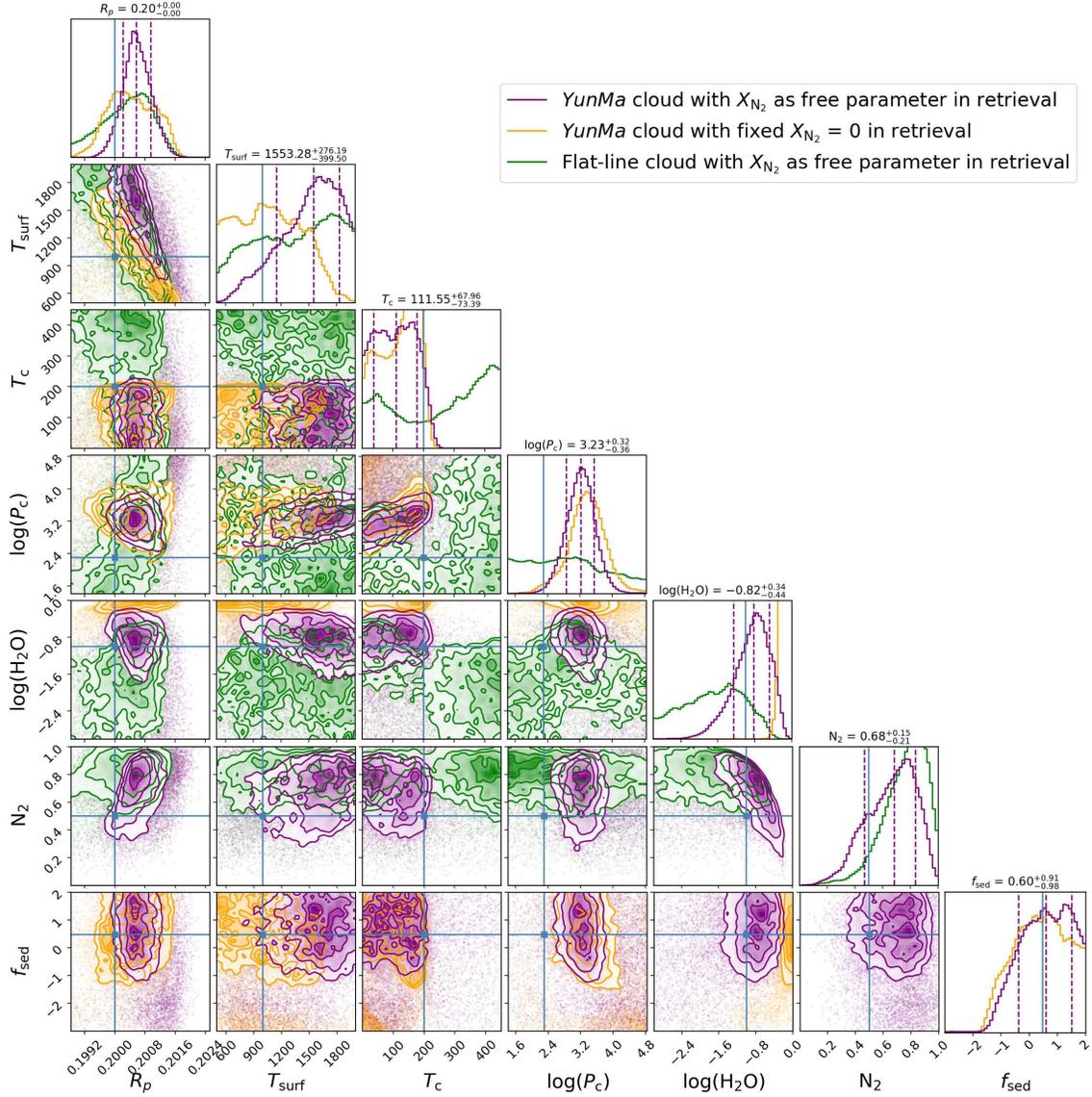}
\caption{Retrieval posteriors for Case \ref{case:p200N205Tp} (purple), \ref{case:p200N205fix0} (orange) and \ref{case:p200N205simple} (green) in Table \ref{tab:retrieval}. Blue crosses indicate the ground truth parameters and the vertical dashed lines in histograms indicate the 1$\sigma$ and 2$\sigma$ confidence ranges of the posterior distribution. These three retrievals use the same transit spectrum as input (thin cloud with $X_{\mathrm{N_2}}$ = 0.5  in Fig. \ref{fig:forwardcloud}) but different retrieval assumptions. Case \ref{case:p200N205Tp} (purple):  H$_2$O and N$_2$ are included as priors and the cloud formation is simulated by \emph{YunMa}. Case \ref{case:p200N205fix0} (orange): same as Case \ref{case:p200N205Tp} except that N$_2$ is not included among the priors. Case \ref{case:p200N205simple} (green): same as Case \ref{case:p200N205Tp} except that the cloud is simulated by a simpler model from \emph{TauREx}, in which the atmosphere becomes opaque below the cloud deck. The only retrieved cloud parameter for the simpler model is the cloud deck pressure, which is not shown here for simplicity. \label{fig:N2simple}}
\end{figure*}

In the experiment of model comparison, Case \ref{case:p200N205simple} simulates the forward spectra with \emph{YunMa} cloud microphysics and then retrieves the atmospheric parameters with another simplified cloud retrieval framework. The simplified clouds are described as an opaque cloud deck across wavelengths in \emph{TauREx 3}, which is commonly used to retrieve data from the last decades with narrow wavelength coverage, e.g., HST/WFC3. In a non-opaque case, we compare the retrieved results from the simple opaque cloud retrieval model in \emph{TauREx 3} (Case \ref{case:p200N205simple}) and \emph{YunMa}. The posteriors using the two different cloud models are compared with each other in Fig. \ref{fig:N2simple}. The former shows relatively flat posterior distributions of the cloud and atmospheric parameters than using \emph{YunMa}, and in this case, lower accuracy and precision of the retrieved results. In Case \ref{case:p200N205nocloud}, we deliberately omit cloud parameters in the retrieval priors and learn if and how other parameters can compensate for those missing. Without the cloud in the prior (Case \ref{case:p200N205nocloud}), the results show a lack of constraints on the $T$-$p$ profile while comparable performance on other parameters with Case \ref{case:p200N205Tp}.

\subsubsection*{Retrievals of featureless spectra}
\begin{figure*}[ht!]
\begin{center}
\plotone{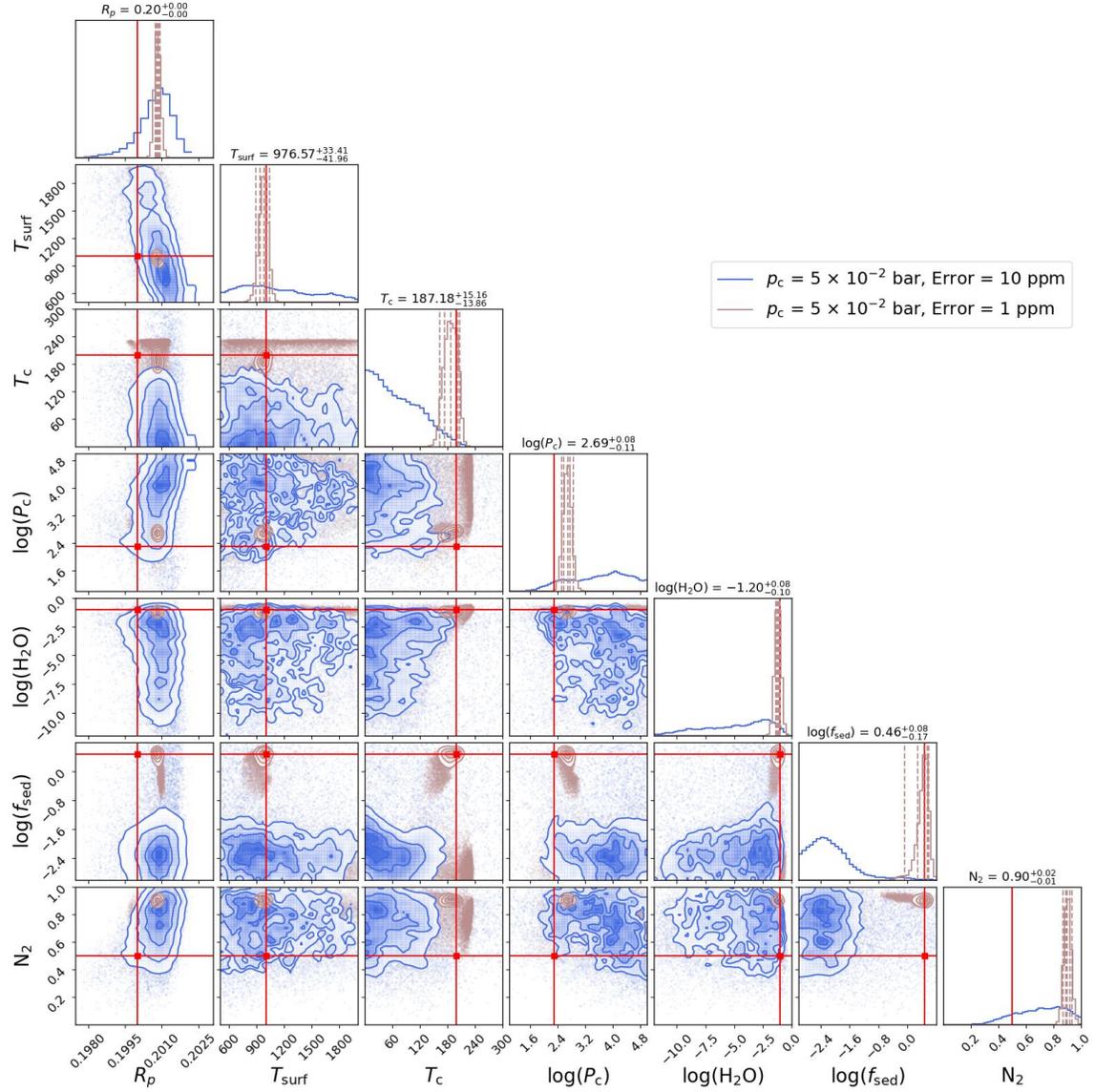}
\caption{\emph{YunMa} retrieval experiments with thick clouds and $X_{\mathrm{N_2}}$ = 0.5  (orange line in Fig. \ref{fig:forwardcloud}). Observational uncertainty = 10 ppm  (blue line, Case \ref{case:p500N205} in Table \ref{tab:retrieval})  and = 1 ppm (brown line, Case \ref{case:p500N2051ppm} in Table \ref{tab:retrieval}). Red lines indicate GTPs. \label{fig:p500N21ppm}}
\end{center}
\end{figure*}

In Cases \ref{case:p500N201}--\ref{case:p500N2051ppm}, we retrieve the atmospheric parameters from spectra with minimal spectroscopic features, where spectral standard deviation ($\sigma_{\mathrm{spec}}$) is less than 1 ppm. The spectra are featureless due to heavy atmospheres and cloud contribution ($p_{\mathrm{c}}$ = 5 $\times$ $10^{-2}$ bar). The transit spectrum for Case \ref{case:p500N205} and \ref{case:p500N2051ppm} is shown in Fig. \ref{fig:forwardcloud}, yellow line: the spectral signal is very small compared to the observational uncertainty (10 ppm). In Case \ref{case:p500N2051ppm}, we unrealistically decrease the uncertainty to 1 ppm at resolving power 100 to evaluate \emph{YunMa}'s performance with idealised data quality. As expected, the retrieval performance greatly improves when the observational uncertainty unrealistically decreases to 1 ppm (Fig. \ref{fig:p500N21ppm}). 

\newcounter{rowcntr}[table]
\renewcommand{\therowcntr}{\arabic{rowcntr}}
\newcolumntype{N}{>{\refstepcounter{rowcntr}\therowcntr}l}

\begin{longrotatetable}
\begin{deluxetable*}{Nlllllll|rrrrrrr}
\tablecaption{\emph{YunMa} retrieval experimental results. The ground truth parameters (GTPs) assumed in the simulations  are listed in the left columns of the table and the retrieved posteriors on the right. The retrieval priors are listed in Table \ref{tab:priors}. In each case, we test the model sensitivity of the atmospheric parameters of $f_{\mathrm{sed}}$, $p_{\mathrm{c}}$, $X_{\mathrm{H_2O}}$, $X_{\mathrm{N_2}}$, and the observational parameters of the error, wavelength coverage ($\lambda$) and the spectral resolution. We choose the nominal value of $p_{\mathrm{c}}$ as 2 $\times$ 10$^{-3}$ bar, where the cloud is not either too thin to be detected or too thick to block the spectroscopic features. \label{tab:retrieval}}
\tablewidth{700pt}
\tabletypesize{\scriptsize}
\tablehead{
\colhead{Case} & \colhead{} & \colhead{} & \colhead{} &\colhead{GTPs} & \colhead{} & \colhead{} & \colhead{} & \colhead{} & \colhead{} & \colhead{} & \colhead{Posteriors} & \colhead{} & \colhead{} & \colhead{}
\\
\hline
\colhead{} & \colhead{log($f_{\mathrm{sed}}$)} & \colhead{log($p_{\mathrm{c}}$)} & \colhead{log($X_{\mathrm{H_2O}}$)} & \colhead{$X_{\mathrm{N_2}}$} & \colhead{Error} & \colhead{$\lambda$} &
\colhead{Res.} &
\colhead{$R_{\mathrm{p}}$} &
\colhead{log($f_{\mathrm{sed}}$)} &
\colhead{log($X_{\mathrm{H_2O}}$)}&
\colhead{$X_{\mathrm{N_2}}$} &
\colhead{log($p_{\mathrm{c}}$)} &
\colhead{$T_{\mathrm{c}}$} &
\colhead{$T_{\mathrm{surf}}$}
\\
\colhead{} & \colhead{} & \colhead{log(bar)} & \colhead{} & \colhead{} & \colhead{(ppm)} & \colhead{(\si{\micro m})} & \colhead{} & \colhead{\nomJovianEqRadius} & \colhead{} & \colhead{} & \colhead{} & \colhead{log(bar)} & \colhead{(K)} & \colhead{(K)}
}
\startdata
\label{case:p100} & 0.48 & -3 & -1 & \nodata & 10 & 0.4 -- 14  & 100 & 0.20$^{+6.98\mathrm{e}-06}_{-4.06\mathrm{e}-06}$ & 0.29$^{+0.00}_{-0.00}$ & -1.10$^{+0.00}_{-0.00}$ & \nodata & \nodata & \nodata & \nodata 
\\ 
\label{case:p200} & 0.48 & -2.7 & -1 & \nodata & 10 & 0.4 -- 14  & 100 & 0.20$^{+1.72\mathrm{e}-03}_{-1.64\mathrm{e}-03}$ & 0.48$^{+0.03}_{-0.02}$ & -0.99$^{+0.17}_{-0.18}$ & \nodata & \nodata & \nodata & \nodata
\\ 
\label{case:p500} & 0.48 & -2.3 & -1 & \nodata & 10 & 0.4 - 14  & 100 & 0.20$^{+3.54\mathrm{e}-03}_{-3.31\mathrm{e}-03}$ & -0.85$^{+0.93}_{-1.56}$ & -1.34$^{+0.47}_{-1.14}$ & \nodata & \nodata & \nodata & \nodata
\\ 
\label{case:p1000} & 0.48 & -2 & -1 & \nodata & 10 & 0.4 -- 14  & 100 & 0.20$^{+5.30\mathrm{e}-03}_{-1.75\mathrm{e}-03}$ & -1.11$^{+0.97}_{-0.75}$ & -1.57$^{+0.79}_{-0.85}$ & \nodata & \nodata & \nodata & \nodata
\\ 
\label{case:p10000} & 0.48 & -1 & -1 & \nodata & 10 & 0.4 -- 14  & 100 & 0.20$^{+1.63\mathrm{e}-03}_{-2.39\mathrm{e}-03}$ & -0.76$^{+0.97}_{-0.81}$ & -1.22$^{+0.31}_{-0.96}$ & \nodata & \nodata & \nodata & \nodata
\\ 
\label{case:p200fsed001} & -2 & -2.7 & -1 & \nodata & 10 & 0.4 -- 14  & 100 & 0.20$^{+2.80\mathrm{e}-03}_{-2.10\mathrm{e}-03}$ & -2.02$^{+0.09}_{-0.11}$ & -0.91$^{+0.28}_{-0.22}$ & \nodata & \nodata & \nodata & \nodata
\\
\label{case:p200fsed10} & 1 & -2.7 & -1 & \nodata & 10 & 0.4 -- 14  & 100 & 0.20$^{+2.12\mathrm{e}-04}_{-7.09\mathrm{e}-04}$ & 1.29$^{+0.43}_{-0.28}$ & -0.92$^{+0.02}_{-0.07}$ & \nodata & \nodata & \nodata & \nodata
\\ 
\label{case:p200H2O001} & 0.48 & -2.7 & -2 & \nodata & 10 & 0.4 -- 14  & 100 & 0.20$^{+2.50\mathrm{e}-05}_{-2.63\mathrm{e}-05}$ & 0.44$^{+0.00}_{-0.01}$ & -2.02$^{+0.01}_{-0.01}$ & \nodata & \nodata & \nodata & \nodata
\\ 
\label{case:p200H2O05} & 0.48 & -2.7 & -0.3 & \nodata & 10 & 0.4 -- 14  & 100 & 0.20$^{+4.43\mathrm{e}-03}_{-3.44\mathrm{e}-03}$ & -1.22$^{+0.95}_{-1.06}$ & -0.75$^{+0.54}_{-0.37}$ & \nodata & \nodata & \nodata & \nodata
\\ 
\label{case:p200err1} & 0.48 & -2.7 & -1 & \nodata & 1 & 0.4 -- 14  & 100 & 0.20$^{+1.62\mathrm{e}-04}_{-1.63\mathrm{e}-04}$ & 0.48$^{+0.00}_{-0.00}$ & -1.00$^{+0.02}_{-0.02}$ & \nodata & \nodata & \nodata & \nodata
\\ 
\label{case:p200err30} & 0.48 & -2.7 & -1 & \nodata & 30 & 0.4 -- 14  & 100 & 0.21$^{+8.61\mathrm{e}-04}_{-3.76\mathrm{e}-03}$ & 0.77$^{+0.85}_{-0.28}$ & -0.50$^{+0.11}_{-0.37}$ & \nodata & \nodata & \nodata & \nodata
\\ 
\label{case:p200res10} & 0.48 & -2.7 & -1 & \nodata & 10 & 0.4 -- 14  & 10 & 0.21$^{+7.33\mathrm{e}-04}_{-2.85\mathrm{e}-03}$ & 0.96$^{+0.70}_{-0.44}$ & -0.47$^{+0.10}_{-0.29}$ & \nodata & \nodata & \nodata & \nodata
\\ 
\label{case:p200wl1} & 0.48 & -2.7 & -1 & \nodata & 10 & 1 -- 14  & 100 & 0.20$^{+2.47\mathrm{e}-03}_{-3.95\mathrm{e}-04}$ & 0.51$^{+0.10}_{-0.01}$ & -0.76$^{+0.25}_{-0.04}$ & \nodata & \nodata & \nodata & \nodata
\\
\label{case:p200Tp} & 0.48 & -2.7 & -1 & \nodata & 10 & 0.4 - 14  & 100 & 0.20$^{+7.53\mathrm{e}-04}_{-9.24\mathrm{e}-04}$ & 0.42$^{+1.04}_{-1.15}$ & -0.62$^{+0.07}_{-0.08}$ & \nodata & -2.39$^{+0.35}_{-0.33}$ & 124.55$^{+56.56}_{-82.38}$ & 1264.49$^{+151.39}_{-121.10}$
\\ 
\label{case:p200Tperr1} & 0.48 & -2.7 & -1 & \nodata & 1 & 0.4 -- 14  & 100 & 0.20$^{+1.39\mathrm{e}-05}_{-1.44\mathrm{e}-05}$ & 0.20$^{+0.01}_{-0.02}$ & -0.71$^{+0.00}_{-0.00}$ & \nodata & -2.54$^{+0.02}_{-0.01}$ & 194.23$^{+1.25}_{-0.88}$ & 1043.45$^{+4.14}_{-1.91}$
\\
\label{case:p500Tp} & 0.48 & -2.3 & -1 & \nodata & 10 & 0.4 -- 14  & 100 & 0.21$^{+7.71\mathrm{e}-04}_{-1.02\mathrm{e}-03}$ & 0.25$^{+1.16}_{-1.24}$ & -0.09$^{+0.06}_{-0.10}$ & \nodata & -1.21$^{+0.52}_{-0.50}$ & 175.25$^{+31.87}_{-86.34}$ & 1237.49$^{+508.48}_{-528.66}$  
\\
\label{case:p1000Tp} & 0.48 & -2 & -1 & \nodata & 10 & 0.4 -- 14  & 100 & 0.20$^{+6.21\mathrm{e}-03}_{-4.08\mathrm{e}-03}$ & -1.84$^{+2.85}_{-0.97}$ & -2.88$^{+2.82}_{-3.98}$ & \nodata & -0.95$^{+0.55}_{-1.03}$ & 150.58$^{+44.47}_{-122.39}$ & 1000.70$^{+706.26}_{-352.40}$
\\
\label{case:p200N2m12} & 0.48 & -2.7 & -1 & 10$^{-12}$ & 10 & 0.4 -- 14  & 100 & 0.21$^{+2.88\mathrm{e}-04}_{-1.72\mathrm{e}-04}$ & 0.62$^{+0.03}_{-0.02}$ & -0.86$^{+0.11}_{-0.12}$ & 0.11$^{+0.03}_{-0.03}$ & \nodata & \nodata & \nodata
\\ 
\label{case:p200N201} & 0.48 & -2.7 & -1 & 0.1 & 10 & 0.4 -- 14  & 100 & 0.20$^{+4.02\mathrm{e}-04}_{-6.63\mathrm{e}-04}$ & 0.67$^{+0.59}_{-0.12}$ & -0.96$^{+0.23}_{-0.24}$ & 0.37$^{+0.09}_{-0.07}$ & \nodata & \nodata & \nodata  
\\
\label{case:p200N205} & 0.48 & -2.7 & -1 & 0.5 & 10 & 0.4 -- 14  & 100 & 0.20$^{+1.82\mathrm{e}-04}_{-2.84\mathrm{e}-04}$ & 0.39$^{+0.12}_{-0.10}$ & -1.03$^{+0.33}_{-0.46}$ & 0.74$^{+0.13}_{-0.13}$ & \nodata & \nodata & \nodata
\\ 
\label{case:p200N205fix05} & 0.48 & -2.7 & -1 & 0.5 & 10 & 0.4 -- 14  & 100 & 0.20$^{+3.96\mathrm{e}-04}_{-4.70\mathrm{e}-04}$ & 0.56$^{+0.96}_{-1.19}$ & -0.69$^{+0.23}_{-0.33}$ & fixed to 0.5 & -1.69$^{+0.32}_{-0.36}$ & 103.11$^{+70.32}_{-69.47}$ & 1443.65$^{+373.77}_{-310.00}$
\\ 
\label{case:p200N205Tp} & 0.48 & -2.7 & -1 & 0.5 & 10 & 0.4 -- 14  & 100 & 0.20$^{+3.87\mathrm{e}-04}_{-3.48\mathrm{e}-04}$ & 0.60$^{+0.91}_{-0.98}$ & -0.82$^{+0.34}_{-0.44}$ & 0.68$^{+0.15}_{-0.21}$ & -1.77$^{+0.32}_{-0.36}$ & 111.53$^{+67.82}_{-73.41}$ & 1553.21$^{+276.23}_{-399.58}$ 
\\ 
\label{case:p200N205fix0} & 0.48 & -2.7 & -1 & 0.5 & 10 & 0.4 -- 14  & 100 & 0.20$^{+7.03\mathrm{e}-04}_{-6.06\mathrm{e}-04}$ & 0.42$^{+1.00}_{-1.02}$ & -0.11$^{+0.07}_{-0.09}$ & fixed to 0 & -1.62$^{+0.42}_{-0.42}$ & 130.80$^{+52.94}_{-86.47}$ & 1071.37$^{+399.26}_{-381.53}$
\\ 
\label{case:p200N205simple} & 0.48 & -2.7 & -1 & 0.5 & 10 & 0.4 -- 14  & 100 & 0.20$^{+6.13\mathrm{e}-04}_{-9.64\mathrm{e}-04}$ & \nodata & -1.90$^{+0.81}_{-1.36}$ & 0.78$^{+0.12}_{-0.18}$ & -2.27$^{+1.22}_{-1.13}$ & 327.32$^{+120.86}_{-245.79}$ & 1370.26$^{+430.51}_{-535.90}$
\\
\label{case:p200N205nocloud} & 0.48 & -2.7 & -1 & 0.5 & 10 & 0.4 -- 14  & 100 & 0.20$^{+3.80\mathrm{e}-04}_{-3.47\mathrm{e}-04}$ & \nodata & -0.95$^{+0.32}_{-0.47}$ & 0.74$^{+0.14}_{-0.16}$ & -1.72$^{+0.38}_{-0.35}$ & 48.60$^{+51.83}_{-32.93}$ & 1577.37$^{+285.82}_{-401.02}$
\\ 
\label{case:p500N201} & 0.48 & -2.3 & -1 & 0.1 & 10 & 0.4 -- 14  & 100 & 0.21$^{+4.38\mathrm{e}-04}_{-8.41\mathrm{e}-04}$ & -2.27$^{+0.67}_{-0.46}$ & -3.80$^{+2.23}_{-3.28}$ & 0.78$^{+0.13}_{-0.18}$ & -1.31$^{+0.82}_{-1.06}$ & 86.15$^{+65.71}_{-61.63}$ & 1136.49$^{+542.96}_{-444.41}$  
\\
\label{case:p500N205} & 0.48 & -2.3 & -1 & 0.5 & 10 & 0.4 -- 14  & 100 & 0.20$^{+4.50\mathrm{e}-04}_{-6.77\mathrm{e}-04}$ & -2.20$^{+0.60}_{-0.48}$ & -4.33$^{+2.36}_{-3.92}$ & 0.72$^{+0.17}_{-0.21}$ & -1.29$^{+0.80}_{-1.11}$ & 62.56$^{+65.77}_{-45.02}$ & 1052.87$^{+540.75}_{-342.32}$  
\\
\label{case:p500N2051ppm} & 0.48 & -2.3 & -1 & 0.5 & 1 & 0.4 -- 14  & 100 & 0.20$^{+4.66\mathrm{e}-05}_{-4.72\mathrm{e}-05}$ & 0.46$^{+0.08}_{-0.17}$ & -1.20$^{+0.08}_{-0.10}$ & 0.90$^{+0.02}_{-0.01}$ & -2.31$^{+0.08}_{-0.11}$ & 187.18$^{+15.16}_{-13.86}$ & 976.55$^{+33.41}_{-41.97}$  
\\
\enddata

\end{deluxetable*}
\end{longrotatetable}


\section{Discussion} \label{sec:discussion}

\subsection{Transit spectroscopy using YunMa}
To understand the performances of \emph{YunMa} in detail, we performed retrieval experiments for over a hundred cases, with different chemistry models, atmospheric and cloud scenarios for super-Earths/sub-Neptunes, hot-Jupiters and brown dwarfs. A variety of cloud species were modelled and analysed. This paper presents a selection of representative examples. We have validated the A-M cloud size distribution in \emph{YunMa} against previous literature simulating NH$_3$ clouds in Jupiter's atmosphere, KCl clouds on artificial large exoplanets and brown dwarfs, and H$_2$O clouds on K2-18 b. We have also tested a number of numerical settings, including fitting methods, tolerances and retrieval samplings. 

In \emph{YunMa}, the mixing ratios of the condensable gas and the condensate are strongly correlated, therefore when cloud forms, the condensable species in the gas phase decreases. Also, a balance is imposed between the upward turbulent mixing and the downward sedimentation velocity from the A-M approach (equation \ref{eq:main}). 
In the current \emph{YunMa}, at each pressure level, the particle number density represents the total number density of particles with different radii. The Earth measurements shown in Fig. 4 of \citet{ackerman2001precipitating} suggest a bimodal distribution of the particle sizes at the same pressure level. \emph{YunMa} is able to use radius bins with their respective number densities to represent more precisely the cloud distribution in the spectral simulation. For the retrieval calculations, however, we had to simplify this information to reduce the computing time.

The cloud opacity is determined by the cloud particle size and number density; different particle sizes have absorption and scattering peaks at different wavelengths. Optically thick clouds cause transit depths with negligible or no modulations as a function of wavelength:
the atmosphere below the cloud deck is, in fact, undetectable while the atmosphere sounded above the cloud deck is more rarefied. Retrieving information about the atmospheric composition and structure is very difficult in the most extreme cases.
For an atmosphere with optically thin clouds, the absorption features due to radiative-active gases (water vapour here) are detectable but less prominent compared with a clear atmosphere. The abundances of these gases can be retrieved largely from their absorption features. If these are condensable species, their abundances further constrain the cloud microphysics. Both the wavelength-dependent features and the overall transit depth help the retrieval performance.

\begin{figure*}[!]
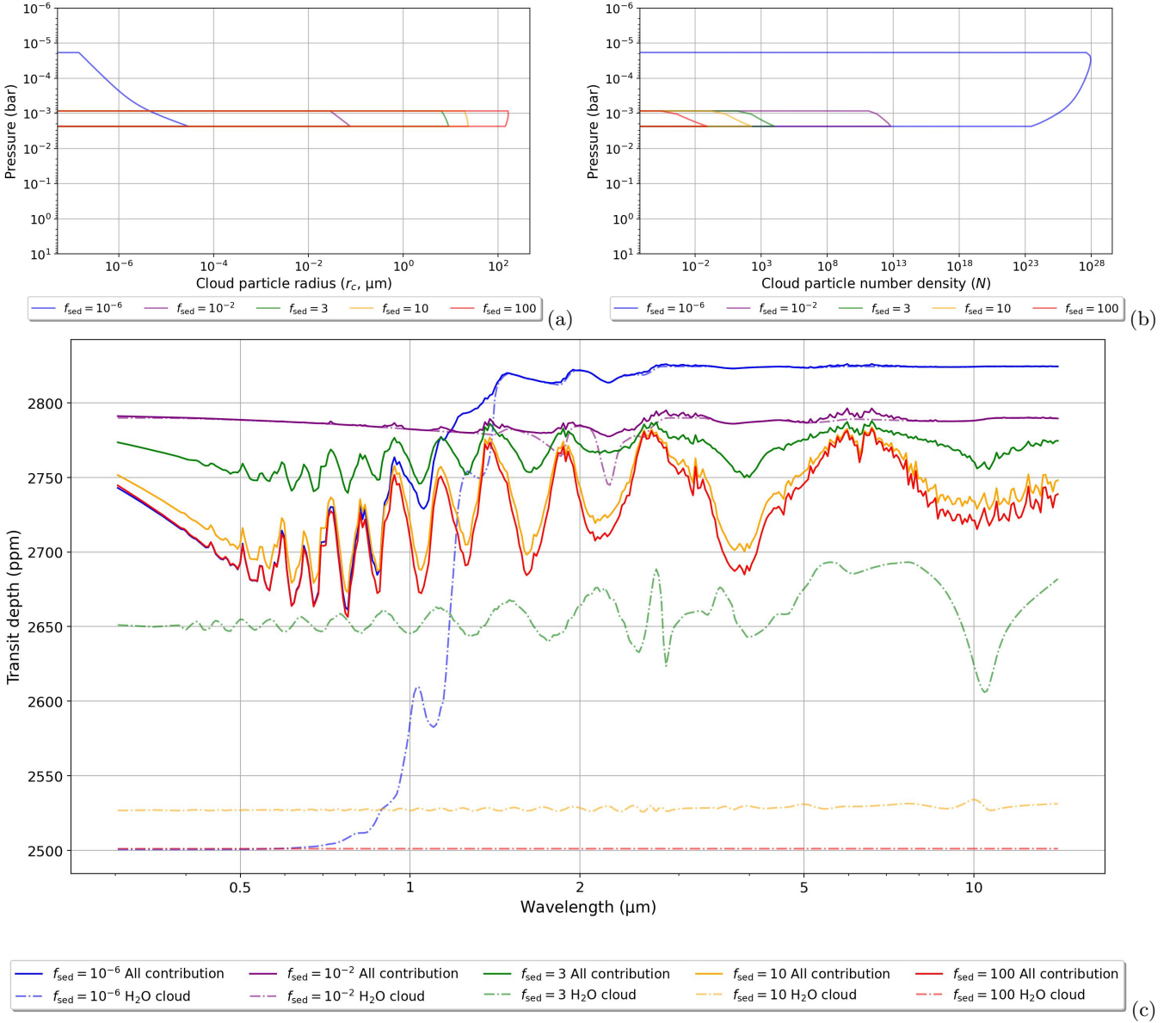

\begin{center}
\includegraphics[width=.47\textwidth]{multifsed_rc.pdf}(a)
\includegraphics[width=.47\textwidth]{multifsed_N.pdf}(b)
\includegraphics[width=.97\textwidth]{multifsed_spec.pdf}(c)
\caption{Test results for $r_\mathrm{c}$ (top left), $N$ (top right) and the transit spectrum (bottom) with different sedimentation efficiency $f_{\mathrm{sed}}$. The cloud layer shrinks with the increasing of $f_{\mathrm{sed}}$.}
\label{fig:fsed_sensitivity}
\end{center}
\end{figure*}

$f_{\mathrm{sed}}$ is the ratio between the sedimentation velocity and the turbulent convective velocity. From the definition, a higher $f_{\mathrm{sed}}$ means a shorter sedimentation timescale, as we fixed $K$ to a constant value. A higher sedimentation efficiency leads to larger offsets to the downward draft, constraining the upward supplement of water vapour and cloud formation. On the contrary, for small $f_{\mathrm{sed}}$, the sedimentation timescale is much longer compared to the diffusion timescale, so the condensation continues at lower pressure to balance the downward sedimentation and upward turbulent mixing; therefore, the cloud region expands.
$f_{\mathrm{sed}}$ is sensitive to the cloud particles' nucleation rate \citep{gao2018sedimentation}. It has a close relationship with the condensate particle size and can be expressed by the particle radius in the lognormal distribution power-law approximation:
\begin{equation}
f_\mathrm{sed}=\frac{\mathrm{\int}_0^\infty r^{3+\alpha}\;\frac{\mathrm{d}n}{\mathrm{d}r}\;\mathrm{d}r}{r_w^\alpha \mathrm{\int}_0^\infty r^3\;\frac{\mathrm{d}n}{\mathrm{d}r}\;\mathrm{d}r},
\label{fsedr}
\end{equation}
which indicates that small $f_{\mathrm{sed}}$ encourages small cloud particle formation. The upward transport is stronger than the sedimentation when the cloud particles are small and vice-versa, which is in line with the experimental results in Figure \ref{fig:fsed_sensitivity} (a, b). The spectrum is sensitive to $f_{\mathrm{sed}}$ when this parameter has values between $10^{-1}$ and $10^{-3}$, as shown in Fig. \ref{fig:fsed_sensitivity} (c), indicating the detectability of $f_{\mathrm{sed}}$ in this interval.

In our experiments, the particle radii are typically 1--10 \si{\micro m}, so they do not block the Rayleigh scattering slope caused by H$_2$ and H$_2$O at the optical wavelengths. For smaller radii (e.g., cyan line in Fig. \ref{fig:residual}), they are expected to contribute more to the optical spectrum, although theoretically it would be hard to form particles at very small sizes according to the nucleation theory in cloud formation \citep{gao2018sedimentation}. For any particle radii even smaller presented in this paper, it is just for model test purposes in extreme cases, and we make no efforts to show their detailed analysis here.

Radiative-inactive gases such as N$_2$, if present, can change the atmospheric scale height. The increase of scale height decreases the transit depth and dampens its spectroscopic features, as illustrated in Fig. \ref{fig:forwardcloud}. As mentioned in previous sections, the presence of radiative-inactive gases cannot be detected directly through spectroscopic signatures. Opaque clouds may behave similarly to inactive gases in mitigating spectroscopic features, leading to potential degeneracy in retrieval experiments. Illustrated in Fig. \ref{fig:N2simple}, the comparison between Case \ref{case:p200N205Tp} and \ref{case:p200N205fix0} in Table \ref{tab:retrieval} suggests the potential degeneracy between $R_{\mathrm{p}}$, the baseline condensable gas abundance, $T$-$p$ profile, and the N$_2$ abundance. In Case \ref{case:p200N205fix0}, we force $X_{\mathrm{N_2}}$ = 0 in the retrieval to monitor how \emph{YunMa} compensates for the missing gases in the atmosphere. Adjustment of $R_{\mathrm{p}}$ translates the transit depth without impact on spectroscopic features. The baseline condensable gas abundance and $T$-$p$ profile are correlated to the formation of clouds. The results indicate that when N$_2$ is absent in prior, the mixing ratio of water vapour -- the radiative-active gas -- is significantly increased to compensate for the missing molecular weight, while more clouds are formed to further reduce the spectroscopic features. The decrease of $T_\mathrm{surf}$ helps in the same way. While the potential degeneracy exists from analysis and the results suggest the model behaviour in the case of missing radiative-inactive gases, the model chose from statistics the scenario closer to the ground truth in Case \ref{case:p200N205Tp}, showing the model's potential in retrieving clouds in heavy atmospheres when it is not opaque. 

As mentioned before, the cloud formation and $T$-$p$ profile are correlated in \emph{YunMa}. Therefore, the presence of optically thin clouds can help to constrain the $T$-$p$ profile in retrievals: for instance, $p_{\mathrm{c}}$, $T_{\mathrm{c}}$ and $T_{\mathrm{surf}}$ are well retrieved in Case \ref{case:p200N205Tp}. On the contrary, the $T$-$p$ profile is not well constrained if clouds are completely absent (Case \ref{case:p200N205nocloud}) or the microphysics part is removed from the retrieval (Case \ref{case:p200N205simple}).

When simulating transit observations, \emph{YunMa} uses a 1-D approach to estimate the cloud formation at terminators according to the thermal profiles present at those locations. If separate observations of the morning and evening terminators are available, these will help us to understand the impact of atmospheric dynamics and horizontal effects. Similarly, phase curves or eclipse ingress and egress observations will be pivotal to completing the 3-D picture of the planet.

\subsection{Cloud formation with the next-generation facilities' data}

With the high-quality transit spectra offered by the next-generation facilities, the uncertainties, wavelength coverage and spectral resolution will be significantly improved compared to most current data, and a simple opaque cloud model is insufficient for the transit study of next-generation data according to our results; the modelling of cloud radiative transfer and microphysics such as \emph{YunMa} is needed. In our experiments, we chose a nominal 10 ppm as the observational uncertainty and found that clouds can be well-characterised in most experiments. With the next-generation data and \emph{YunMa}, there are still limits in retrieving the featureless spectra. A smaller uncertainly may help in retrieving cloud parameters in the most difficult cases: we therefore adopted an unrealistic 1 ppm to test their detectability in an ideal case (Case \ref{case:p200err1}, \ref{case:p200Tperr1} and \ref{fig:p500N21ppm}). With 1 ppm, the atmospheres were constrained well, although the spectra were featureless.  
The results of Case \ref{case:p200} and \ref{case:p200wl1} suggest that broad wavelength coverage is paramount to characterising clouds well: here, optical wavelengths play a critical role when combined with infrared spectral coverage.

\subsection{Numerical instabilities in cloud microphysics simulations}
\begin{figure*}[ht!]
\plotone{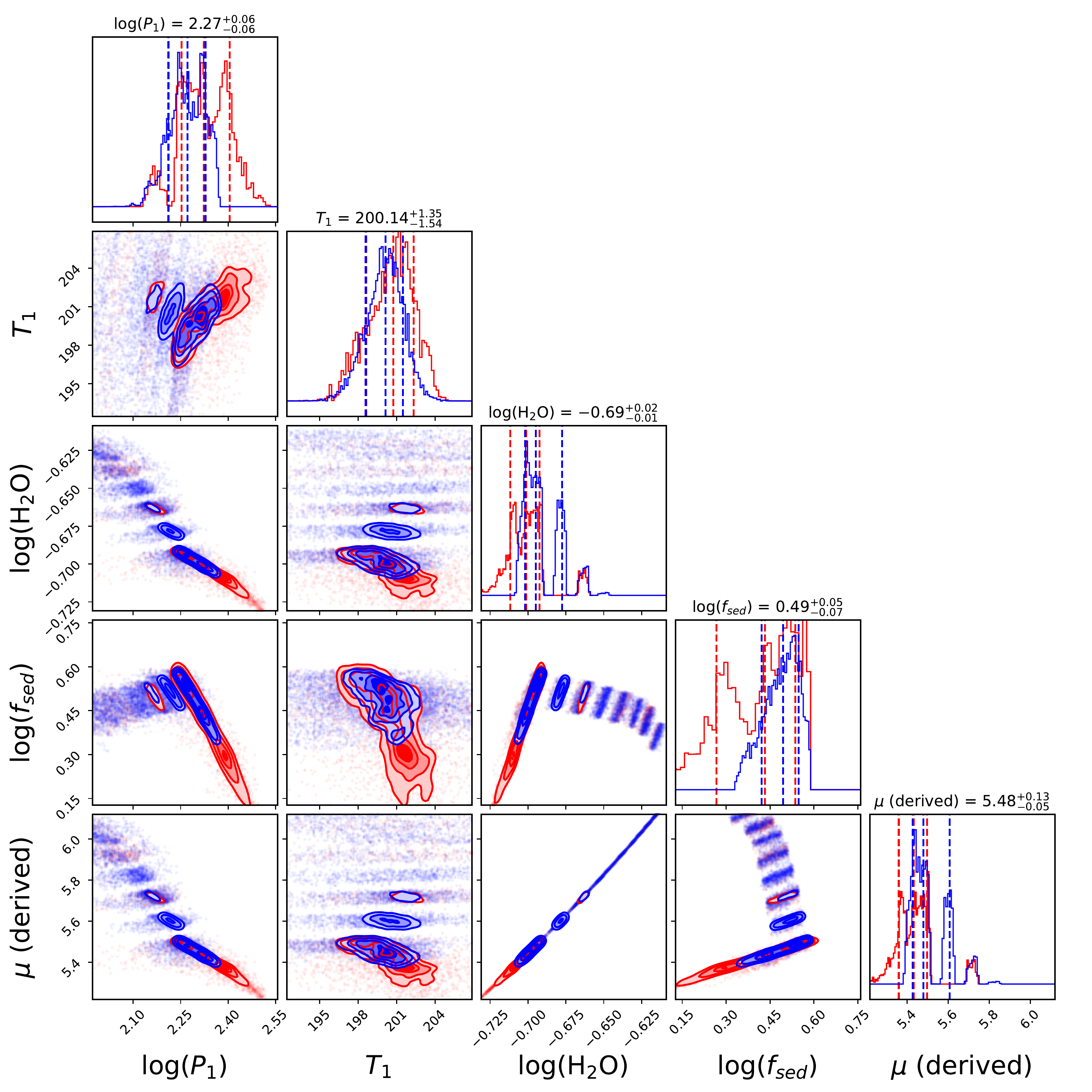}
\caption{Example of numerical instability in cloud formation simulations when large relative ($rtol$) and absolute ($atol$) tolerances are chosen in solving equation (\ref{eq:main}) with Runge-Kutta method. Blue plots: $rtol$ = 1 $\times$ 10$^{-8}$ and $atol$ = 1 $\times$ 10$^{-12}$. Red plots:  $rtol$ = 1 $\times$ 10$^{-12}$ and $atol$ = 1 $\times$ 10$^{-15}$. In these tests, the retrieval performances are affected by numerical instability, as illustrated in this figure where multiple islands of solutions in the posteriors are clearly visible. \label{fig:numerical}}
\end{figure*}

We have selected the explicit Runge-Kutta method of order 8 \citep[DOP853,][]{Hairer1988} to solve equation (\ref{eq:main}) as it delivered the most stable numerical performance for our experiments. However, numerical instabilities may occur when a large relative ($rtol$) and absolute ($atol$) tolerances are chosen. Sometimes the ODE solver cannot converge for $q_\mathrm{t}$ and indicates ``no clouds'' as a solution due to the numerical instability. Caution is also needed in estimating the cloud mixing ratio ($q_\mathrm{c}$), as $q_\mathrm{t}$ might be two orders of magnitude larger than $q_\mathrm{c}$ and numerical errors could be injected when $q_s$ is subtracted from $q_\mathrm{t}$. When solving the ODE with too large tolerance, $q_\mathrm{t}$ might converge to a certain value, but $q_\mathrm{c}$ would be estimated as negligible according to equation (\ref{eq:main} and \ref{eq:qc_condense}). In other words, even though when the ODE solution for the gas phase seems numerically stable, it might not be precise and accurate enough. In those cases, the retrieval performances are affected, as shown in Fig. \ref{fig:numerical}, where multiple islands of solutions in the posteriors are visible; these are caused by numerical instabilities, and the issue is more obvious for larger tolerance values. After many tests, we have decided to use ``DOP853'' in solving the ODE with $rtol$ = 1 $\times$ $10^{-13}$ and $atol$ = $10^{-16}$, which guarantee numerically stable results.

\section{Conclusions}
\emph{YunMa} is a state-of-art cloud simulation and retrieval package optimised for the interpretation of the next generation of exoplanetary atmospheric data, as provided by e.g., JWST, Roman, Twinkle and Ariel. These facilities will provide an unprecedented amount of high-quality data, where the cloud formation process and cloud scattering properties can no longer be ignored. 

\emph{YunMa} cloud microphysics is based on the model published by  \cite{ackerman2001precipitating}, while the scattering properties of clouds are calculated through the
open-source BH-Mie code. 
When coupled to the \emph{TauREx} framework \citep{alrefaie2021taurex31}, \emph{YunMa} becomes a very versatile model that can simulate transit and eclipse spectra for a variety of cloudy exoplanets with different masses, atmospheric compositions and temperatures.
Most importantly, \emph{YunMa}+\emph{TauREx} can be used as a spectral retrieval framework optimised for cloudy atmospheres. 

We have validated \emph{YunMa} against previous work that adopted the A-M approach and compared our results with a 3-D model simulation when consistent assumptions are adopted to produce the vertical profile. We have validated the radiative transfer calculations in \emph{YunMa}, including cloud scattering, against PyMieScatt. 

We have run over one hundred retrieval experiments with \emph{YunMa}, with different cloud compositions (e.g., KCl, MgSiO$_3$, Fe clouds). This paper presents and discusses 28 cases of water clouds in the atmosphere of a temperate super-Earth, K2-18 b-like. 
Through these experiments, we have learnt that \emph{YunMa} is capable of retrieving cloud formation and atmospheric parameters when clouds are not so opaque to mask all the atmospheric features at most wavelengths. More specifically, if we assume spectroscopic data covering the 0.4--14 \si{\micro m} range, uncertainties at a level of 10 ppm and spectral resolution R = 100, we can retrieve the sedimentation coefficient, the baseline condensable gases and the $T$-$p$ profile points with accuracy levels $>$ 60$\%$ in our respective experiments. This is not the case using the simple opaque cloud model, which shows more degeneracy in the posterior distributions of the atmospheric parameters retrieved.

An extension of \emph{YunMa} to interpret phase-curve observations will be a valuable next step. 2-D cloud models, which include horizontal convection, will soon be within reach of computing speed and might be considered in future versions of \emph{YunMa}. While we are aware of the limitations of the specific cloud microphysics model embedded in \emph{YunMa}, our results advocate for the need to include more realistic cloud models in spectral retrievals to interpret correctly the results of the next-generation facilities.



\appendix

\section{Supplementary equations to estimate the cloud mixing profile} \label{appendix:eqA}
In Section \ref{subsection:qc}, the SVP can be estimated with the Clausius-Clapeyron equation:
\begin{equation}
\label{eq:sat}
e_s = e_0 \; \mathrm{exp}\left[ \frac{\ell}{R_{\mathrm{SV}}}\left(\frac{1}{T_0}-\frac{1}{T}\right)\right],
\end{equation}
where $\ell$ is the latent heat of evaporation, $R_{\mathrm{SV}}$ is the specific gas constant for the vapour, $T$ is the atmospheric temperature, and $T_0$ is the temperature at vapour pressure $e_0$. We use the results from laboratory measurements of these parameters for the different chemical species when available.

$\beta$ is the Cunningham slip factor:
\begin{equation}
\label{eq:beta}
\beta = 1 + 1.26N_{\mathrm{Kn}},
\end{equation}
The Knudsen number ($N_{\mathrm{Kn}}$) is the ratio between the molecular mean free path and the droplet radius.

\emph{YunMa} adopts two ways to estimate $\eta$. One is that \cite{lavvas2008coupling} suggested to use:
\begin{equation}
\label{eq:lavva}
    \eta = \frac{1}{3} \rho_\mathrm{a} \overline{V} \lambda_\mathrm{a},
\end{equation}
where $\overline{V}$ is the thermal velocity of gaseous components, and $\lambda_\mathrm{a}$ is the mean free path. Another uses the definition by \citet{rosner2012transport}, which is also adopted by \citet{ackerman2001precipitating}:
\begin{equation}
\label{eq:rosner}
\eta = \frac{5}{16} \frac{\sqrt{\pi m k_{\mathrm{B}} T}}{\pi d^2} \frac{(k_{\mathrm{B}}T / \epsilon)^{0.16}}{1.22},
\end{equation}
where $d$ is the diameter of a gas particle and $\epsilon$ is the atmospheric Lennard-Jones potential well depth. When using the A-M approach, $v_f$ and the particle size are positively correlated using $\eta$ either from \cite{rosner2012transport} or \cite{lavvas2008coupling}.

\section{Derivation of the cloud particle size and number density} \label{appendix:eqB}
In Section \ref{subsection:r}, assuming a lognormal cloud particle size distribution, the geometric mean ($r_\mathrm{g}$) is defined as: 
\begin{equation}
\label{eq:rg}
r_\mathrm{g} = e^{\frac{\int_0^\infty \mathrm{ln} r \;\frac{\mathrm{d}n}{\mathrm{d}r}\; \mathrm{d}r}{\int_0^\infty \;\frac{\mathrm{d}n}{\mathrm{d}r}\; \mathrm{d}r}}.
\end{equation}
\noindent The power law approximation allows representation of $f_\mathrm{sed}$ using the particle size distribution. 

\begin{equation}
\label{eq:fsed2}
f_{\mathrm{sed}} \approx \frac{\mathrm{\int}_0^\infty r^{3+\alpha} \; \frac{\mathrm{d}n}{\mathrm{d}r} \; \mathrm{d}r}{r_w^\alpha\mathrm{\int}_0^\infty \; r^{3}\frac{\mathrm{d}n}{\mathrm{d}r} \; \mathrm{d}r},
\end{equation}
where $n$ is the accumulated number density as defined in Section \ref{sec:model} and $\sigma_\mathrm{g}$ is the geometric standard deviation of the lognormal particle radius distribution.
\noindent Through an integration of the lognormal distribution, \citet{ackerman2001precipitating} derived that:
\begin{equation}
r_\mathrm{g} = r_w \; f_{\mathrm{sed}}^{\frac{1}{\alpha}}\; \mathrm{exp} \left(-\frac{\alpha + 6}{2}\; \mathrm{ln}^2 \sigma_\mathrm{g} \right ),
\end{equation}
where the $\sigma_\mathrm{g}$ is the geometric standard deviation of the particle radii. 

The effective mean radius ($r_{\mathrm{eff}}$) is the area-weighted average radius defined by \citet{Hansen1974eff} to approximately represent the scattering properties of the whole size distribution by a single parameter when the particle radius is larger than the radiation wavelength. To derive $r_{\mathrm{eff}}$, we first recall that in lognormal distribution, the $t$-th raw moment is given by:
\begin{equation}
\label{eq:moment}
m_t = N_0 r_\mathrm{g}\;\mathrm{exp}\left[\frac{(t\sigma)^2}{2}\right].
\end{equation}

\noindent Therefore $r_\mathrm{eff}$, which is the area-weighted average radius, can be estimated through:
\begin{equation}
\label{eq:reff}
r_{\mathrm{eff}} = \frac{\mathrm{\int}_0^\infty r\pi r^2 \;\frac{\mathrm{d}n}{\mathrm{d}r}\; \mathrm{d}r}{\mathrm{\int}_0^\infty \pi r^2 \;\frac{\mathrm{d}n}{\mathrm{d}r}\; \mathrm{d}r} = r_w \; f_{\mathrm{sed}}^{\frac{1}{\alpha}}\; \mathrm{exp} \left(-\frac{\alpha + 1}{2}\; \mathrm{ln^2} \sigma_\mathrm{g} \right).
\end{equation}

Similarly, the total number density for the particles is estimated by using the volume-weighted mean: 
\begin{equation}
N = \frac{3 \varepsilon \rho_\mathrm{a} q_\mathrm{c}}{4 \pi \rho_\mathrm{p} r_\mathrm{g}^3} \ \mathrm{exp} \left(-\frac{9}{2}\ \mathrm{ln^2} \sigma_\mathrm{g} \right).
\end{equation}

\begin{acknowledgments}
The work presented in this paper was partially supported by UKSA, grant ST/X002616/1 and ST/W00254X/1, and ExoMolHD ERC grant 883830. YI is supported by JSPS KAKENHI JP22K14090. QC is supported by the ESA Research Fellowship Program. The authors acknowledge the use of the UCL Kathleen High-Performance Computing Facility (Kathleen@UCL) and associated support services in completing this work. This work utilised the Cambridge Service for Data-Driven Discovery (CSD3), part of which is operated by the University of Cambridge Research Computing on behalf of the STFC DiRAC HPC Facility (dirac.ac.uk). The DiRAC component of CSD3 was funded by BEIS capital funding via STFC capital grants ST/P002307/1 and ST/R002452/1 and STFC operations grant ST/R00689X/1. DiRAC is part of the National e-Infrastructure.

We thank Dr Andrew S. Ackerman for the constructive communication on the A-M model and for sharing the original code, Dr Kai-Hou Yip and Sam Wright for suggestions on the numerical methods and Dr Yui Kawashima for the cloud and haze model discussion. We thank the anonymous reviewer for the constructive comments, which improved the paper greatly.
\end{acknowledgments}

\bibliography{YunMa_Enabling_Spectral_Retrievals_of_Exoplanetary_Clouds}{}

\begin{thebibliography}{}
\expandafter\ifx\csname natexlab\endcsname\relax\def\natexlab#1{#1}\fi
\providecommand{\url}[1]{\href{#1}{#1}}
\providecommand{\dodoi}[1]{doi:~\href{http://doi.org/#1}{\nolinkurl{#1}}}
\providecommand{\doeprint}[1]{\href{http://ascl.net/#1}{\nolinkurl{http://ascl.net/#1}}}
\providecommand{\doarXiv}[1]{\href{https://arxiv.org/abs/#1}{\nolinkurl{https://arxiv.org/abs/#1}}}

\bibitem[{Ackerman \& Marley(2001)}]{ackerman2001precipitating}
Ackerman, A.~S., \& Marley, M.~S. 2001, \apj, 556, 872, \dodoi{10.1086/321540}

\bibitem[{{Adams} {et~al.}(2022){Adams}, {Kataria}, {Batalha}, {Gao}, \& {Knutson}}]{adams2022hotjupiters}
{Adams}, D.~J., {Kataria}, T., {Batalha}, N.~E., {Gao}, P., \& {Knutson}, H.~A. 2022, \apj, 926, 157, \dodoi{10.3847/1538-4357/ac3d32}

\bibitem[{Al-Refaie {et~al.}(2022{\natexlab{a}})Al-Refaie, Changeat, Venot, Waldmann, \& Tinetti}]{Al-Refaie2022taurex}
Al-Refaie, A.~F., Changeat, Q., Venot, O., Waldmann, I.~P., \& Tinetti, G. 2022{\natexlab{a}}, \apj, 932, 123, \dodoi{10.3847/1538-4357/ac6dcd}

\bibitem[{{Al-Refaie} {et~al.}(2021){Al-Refaie}, Changeat, Waldmann, \& Tinetti}]{alrefaie2021taurex31}
{Al-Refaie}, A.~F., Changeat, Q., Waldmann, I.~P., \& Tinetti, G. 2021, \apj, 917, 37, \dodoi{10.3847/1538-4357/ac0252}

\bibitem[{Al-Refaie {et~al.}(2022{\natexlab{b}})Al-Refaie, Venot, Changeat, \& Edwards}]{alrefaie2022freckll}
Al-Refaie, A.~F., Venot, O., Changeat, Q., \& Edwards, B. 2022{\natexlab{b}}.
\newblock \doarXiv{2209.11203}

\bibitem[{Baeyens {et~al.}(2021)Baeyens, Decin, Carone, Venot, AgÃºndez, \& MolliÃšre}]{Baeyens2021}
Baeyens, R., Decin, L., Carone, L., {et~al.} 2021, \mnras, 505, 5603, \dodoi{10.1093/mnras/stab1310}

\bibitem[{{Barstow}(2020)}]{barstow2020parametric}
{Barstow}, J.~K. 2020, \mnras, 497, 4183, \dodoi{10.1093/mnras/staa2219}

\bibitem[{{Batalha} {et~al.}(2019){Batalha}, {Marley}, {Lewis}, \& {Fortney}}]{Batalha2019picaso}
{Batalha}, N.~E., {Marley}, M.~S., {Lewis}, N.~K., \& {Fortney}, J.~J. 2019, \apj, 878, 70, \dodoi{10.3847/1538-4357/ab1b51}

\bibitem[{{Baudino} {et~al.}(2015){Baudino}, {B{\'e}zard}, {Boccaletti}, {Bonnefoy}, {Lagrange}, \& {Galicher}}]{baudino2015exorem}
{Baudino}, J.~L., {B{\'e}zard}, B., {Boccaletti}, A., {et~al.} 2015, \aap, 582, A83, \dodoi{10.1051/0004-6361/201526332}

\bibitem[{{Bean} {et~al.}(2018){Bean}, {Stevenson}, {Batalha}, {Berta-Thompson}, {Kreidberg}, {Crouzet}, {Benneke}, {Line}, {Sing}, {Wakeford}, {Knutson}, {Kempton}, {D{\'e}sert}, {Crossfield}, {Batalha}, {de Wit}, {Parmentier}, {Harrington}, {Moses}, {Lopez-Morales}, {Alam}, {Blecic}, {Bruno}, {Carter}, {Chapman}, {Decin}, {Dragomir}, {Evans}, {Fortney}, {Fraine}, {Gao}, {Garc{\'\i}a Mu{\~n}oz}, {Gibson}, {Goyal}, {Heng}, {Hu}, {Kendrew}, {Kilpatrick}, {Krick}, {Lagage}, {Lendl}, {Louden}, {Madhusudhan}, {Mandell}, {Mansfield}, {May}, {Morello}, {Morley}, {Nikolov}, {Redfield}, {Roberts}, {Schlawin}, {Spake}, {Todorov}, {Tsiaras}, {Venot}, {Waalkes}, {Wheatley}, {Zellem}, {Angerhausen}, {Barrado}, {Carone}, {Casewell}, {Cubillos}, {Damiano}, {de Val-Borro}, {Drummond}, {Edwards}, {Endl}, {Espinoza}, {France}, {Gizis}, {Greene}, {Henning}, {Hong}, {Ingalls}, {Iro}, {Irwin}, {Kataria}, {Lahuis}, {Leconte}, {Lillo-Box}, {Lines}, {Lothringer}, {Mancini}, {Marchis}, {Mayne}, {Palle}, {Rauscher}, {Roudier},
  {Shkolnik}, {Southworth}, {Swain}, {Taylor}, {Teske}, {Tinetti}, {Tremblin}, {Tucker}, {van Boekel}, {Waldmann}, {Weaver}, \& {Zingales}}]{Bean2018jwst}
{Bean}, J.~L., {Stevenson}, K.~B., {Batalha}, N.~M., {et~al.} 2018, \pasp, 130, 114402, \dodoi{10.1088/1538-3873/aadbf3}

\bibitem[{{Benneke}(2015)}]{Benneke2015Scarlet}
{Benneke}, B. 2015.
\newblock \doarXiv{1504.07655}

\bibitem[{Benneke {et~al.}(2019 a)Benneke, Knutson, Lothringer, Crossfield, Moses, Morley, Kreidberg, Fulton, Dragomir, Howard, {et~al.}}]{benneke2019sub}
Benneke, B., Knutson, H.~A., Lothringer, J., {et~al.} 2019 a, Nature Astronomy, 3, 813, \dodoi{10.1038/s41550-019-0800-5}

\bibitem[{Bohren \& Huffman(2008)}]{bohren2008absorption}
Bohren, C.~F., \& Huffman, D.~R. 2008, Absorption and scattering of light by small particles (New York: John Wiley \& Sons)

\bibitem[{{Boucher} {et~al.}(2021){Boucher}, {Darveau-Bernier}, {Pelletier}, {Lafreni{\`e}re}, {Artigau}, {Cook}, {Allart}, {Radica}, {Doyon}, {Benneke}, {Arnold}, {Bonfils}, {Bourrier}, {Cloutier}, {Gomes da Silva}, {Deibert}, {Delfosse}, {Donati}, {Ehrenreich}, {Figueira}, {Forveille}, {Fouqu{\'e}}, {Gagn{\'e}}, {Gaidos}, {H{\'e}brard}, {Jayawardhana}, {Klein}, {Lovis}, {Martins}, {Martioli}, {Moutou}, \& {Santos}}]{boucher2021gray}
{Boucher}, A., {Darveau-Bernier}, A., {Pelletier}, S., {et~al.} 2021, \aj, 162, 233, \dodoi{10.3847/1538-3881/ac1f8e}

\bibitem[{{Brogi} \& {Line}(2019)}]{brogi2019grey}
{Brogi}, M., \& {Line}, M.~R. 2019, \aj, 157, 114, \dodoi{10.3847/1538-3881/aaffd3}

\bibitem[{{Burrows}(2014)}]{Burrows2014review}
{Burrows}, A.~S. 2014, \nat, 513, 345, \dodoi{10.1038/nature13782}

\bibitem[{{Caldas} {et~al.}(2019){Caldas}, {Leconte}, {Selsis}, {Waldmann}, {Bord{\'e}}, {Rocchetto}, \& {Charnay}}]{Caldas2019bias}
{Caldas}, A., {Leconte}, J., {Selsis}, F., {et~al.} 2019, \aap, 623, A161, \dodoi{10.1051/0004-6361/201834384}

\bibitem[{Carlson {et~al.}(1988)Carlson, Rossow, \& Orton}]{carlson1988cloud}
Carlson, B.~E., Rossow, W.~B., \& Orton, G.~S. 1988, Journal of Atmospheric Sciences, 45, 2066, \dodoi{10.1175/1520-0469(1988)045<2066:CMOTGP>2.0.CO;2}

\bibitem[{{Changeat} {et~al.}(2021{\natexlab{a}}){Changeat}, {Al-Refaie}, {Edwards}, {Waldmann}, \& {Tinetti}}]{changeat2021bias}
{Changeat}, Q., {Al-Refaie}, A.~F., {Edwards}, B., {Waldmann}, I.~P., \& {Tinetti}, G. 2021{\natexlab{a}}, \apj, 913, 73, \dodoi{10.3847/1538-4357/abf2bb}

\bibitem[{{Changeat} {et~al.}(2021{\natexlab{b}}){Changeat}, {Edwards}, {Al-Refaie}, {Tsiaras}, {Waldmann}, \& {Tinetti}}]{changeat2021disentangling}
{Changeat}, Q., {Edwards}, B., {Al-Refaie}, A.~F., {et~al.} 2021{\natexlab{b}}, Experimental Astronomy, \dodoi{10.1007/s10686-021-09794-w}

\bibitem[{{{Changeat}} {et~al.}(2022){{Changeat}}, {Edwards}, {Al-Refaie}, {Tsiaras}, {Skinner}, {Cho}, {Yip}, {Anisman}, {Ikoma}, {Bieger}, {Venot}, {Shibata}, {Waldmann}, \& {Tinetti}}]{changeat2022five}
{{Changeat}}, Q., {Edwards}, B., {Al-Refaie}, A.~F., {et~al.} 2022, \apjs, 260, 3, \dodoi{10.3847/1538-4365/ac5cc2}

\bibitem[{{Charnay} {et~al.}(2018){Charnay}, {B{\'e}zard}, {Baudino}, {Bonnefoy}, {Boccaletti}, \& {Galicher}}]{charnay2018exorem}
{Charnay}, B., {B{\'e}zard}, B., {Baudino}, J.~L., {et~al.} 2018, \apj, 854, 172, \dodoi{10.3847/1538-4357/aaac7d}

\bibitem[{Charnay {et~al.}(2021 b)Charnay, Blain, B{\'e}zard, Leconte, Turbet, \& Falco}]{charnay2021formation}
Charnay, B., Blain, D., B{\'e}zard, B., {et~al.} 2021 b, \aap, 646, A171, \dodoi{10.1051/0004-6361/202039525}

\bibitem[{{Charnay} {et~al.}(2022){Charnay}, {Tobie}, {Lebonnois}, \& {Lorenz}}]{charnay2022titan}
{Charnay}, B., {Tobie}, G., {Lebonnois}, S., \& {Lorenz}, R.~D. 2022, \aap, 658, A108, \dodoi{10.1051/0004-6361/202141898}

\bibitem[{Cho {et~al.}(2021)Cho, Skinner, \& Thrastarson}]{Cho_2021}
Cho, J. Y.-K., Skinner, J.~W., \& Thrastarson, H.~T. 2021, \apjl, 913, L32, \dodoi{10.3847/2041-8213/abfd37}

\bibitem[{{Chubb} {et~al.}(2021){Chubb}, {Rocchetto}, {Yurchenko}, {Min}, {Waldmann}, {Barstow}, {Molli{\`e}re}, {Al-Refaie}, {Phillips}, \& {Tennyson}}]{chubb2022exomol}
{Chubb}, K.~L., {Rocchetto}, M., {Yurchenko}, S.~N., {et~al.} 2021, \aap, 646, A21, \dodoi{10.1051/0004-6361/202038350}

\bibitem[{Cox(2015)}]{Cox2015allen}
Cox, A.~N. 2015, Allen’s Astrophysical Quantities (New York: Springer)

\bibitem[{{Cubillos} \& {Blecic}(2021)}]{Cubillos2021PYRAT}
{Cubillos}, P.~E., \& {Blecic}, J. 2021, \mnras, 505, 2675, \dodoi{10.1093/mnras/stab1405}

\bibitem[{{Drummond, B.} {et~al.}(2016){Drummond, B.}, {Tremblin, P.}, {Baraffe, I.}, {Amundsen, D. S.}, {Mayne, N. J.}, {Venot, O.}, \& {Goyal, J.}}]{Drummond2016ATMO}
{Drummond, B.}, {Tremblin, P.}, {Baraffe, I.}, {et~al.} 2016, A\&A, 594, A69, \dodoi{10.1051/0004-6361/201628799}

\bibitem[{Edwards {et~al.}(2019)Edwards, Rice, Zingales, Tessenyi, Waldmann, Tinetti, Pascale, Savini, \& Sarkar}]{edwards2019exoplanet}
Edwards, B., Rice, M., Zingales, T., {et~al.} 2019, Experimental Astronomy, 47, 29, \dodoi{10.1007/s10686-018-9611-4}

\bibitem[{{Edwards} {et~al.}(2022){Edwards}, {Changeat}, {Tsiaras}, {Hou Yip}, {Al-Refaie}, {Anisman}, {Bieger}, {Gressier}, {Shibata}, {Skaf}, {Bouwman}, {Y-K. Cho}, {Ikoma}, {Venot}, {Waldmann}, {Lagage}, \& {Tinetti}}]{Edwards2022G141}
{Edwards}, B., {Changeat}, Q., {Tsiaras}, A., {et~al.} 2022.
\newblock \doarXiv{2211.00649}

\bibitem[{{Fleck} \& {Canfield}(1984)}]{mie1984scattering}
{Fleck}, J.~A., J., \& {Canfield}, E.~H. 1984, Journal of Computational Physics, 54, 508, \dodoi{10.1016/0021-9991(84)90130-X}

\bibitem[{{Forget} {et~al.}(1999){Forget}, {Hourdin}, {Fournier}, {Hourdin}, {Talagrand}, {Collins}, {Lewis}, {Read}, \& {Huot}}]{Forget1999Mars}
{Forget}, F., {Hourdin}, F., {Fournier}, R., {et~al.} 1999, \jgr, 104, 24155, \dodoi{10.1029/1999JE001025}

\bibitem[{{Fortney} {et~al.}(2021){Fortney}, {Barstow}, \& {Madhusudhan}}]{Fortney2021review}
{Fortney}, J.~J., {Barstow}, J.~K., \& {Madhusudhan}, N. 2021, in ExoFrontiers; Big Questions in Exoplanetary Science, ed. N.~{Madhusudhan} (Bristol: IOP Publishing), 17--1, \dodoi{10.1088/2514-3433/abfa8fch17}

\bibitem[{{Gandhi} \& {Madhusudhan}(2018)}]{Gandhi2018HyDRA}
{Gandhi}, S., \& {Madhusudhan}, N. 2018, \mnras, 474, 271, \dodoi{10.1093/mnras/stx2748}

\bibitem[{Gao {et~al.}(2018)Gao, Marley, \& Ackerman}]{gao2018sedimentation}
Gao, P., Marley, M.~S., \& Ackerman, A.~S. 2018, \apj, 855, 86, \dodoi{10.3847/1538-4357/aab0a1}

\bibitem[{{Gao} {et~al.}(2020){Gao}, {Thorngren}, {Lee}, {Fortney}, {Morley}, {Wakeford}, {Powell}, {Stevenson}, \& {Zhang}}]{Gao2020cloudspec}
{Gao}, P., {Thorngren}, D.~P., {Lee}, E. K.~H., {et~al.} 2020, Nature Astronomy, 4, 951, \dodoi{10.1038/s41550-020-1114-3}

\bibitem[{Gardner {et~al.}(2006)Gardner, Mather, Clampin, Doyon, Greenhouse, Hammel, Hutchings, Jakobsen, Lilly, Long, {et~al.}}]{gardner2006james}
Gardner, J.~P., Mather, J.~C., Clampin, M., {et~al.} 2006, Space Science Reviews, 123, 485, \dodoi{10.1007/s11214-006-8315-7}

\bibitem[{{Gierasch} \& {Conrath}(1985)}]{gierasch1985energy}
{Gierasch}, P.~J., \& {Conrath}, B.~J. 1985, in Recent Advances in Planetary Meteorology, ed. G.~E. {Hunt} (Cambridge: Cambridge Univ. Press), 121--146

\bibitem[{Goyal {et~al.}(2017)Goyal, Mayne, Sing, Drummond, Tremblin, Amundsen, Evans, Carter, Spake, Baraffe, Nikolov, Manners, Chabrier, \& Hebrard}]{Goyal2017ATMO}
Goyal, J.~M., Mayne, N., Sing, D.~K., {et~al.} 2017, \mnras, 474, 5158, \dodoi{10.1093/mnras/stx3015}

\bibitem[{{Greene} {et~al.}(2016){Greene}, {Line}, {Montero}, {Fortney}, {Lustig-Yaeger}, \& {Luther}}]{Greene2016jwst}
{Greene}, T.~P., {Line}, M.~R., {Montero}, C., {et~al.} 2016, \apj, 817, 17, \dodoi{10.3847/0004-637X/817/1/17}

\bibitem[{{Hairer} {et~al.}(1993){Hairer}, {Norsett}, \& {Wanner}}]{Hairer1988}
{Hairer}, E., {Norsett}, S.~P., \& {Wanner}, G. 1993, Solving Ordinary Differential Equations. I: Nonstiff Problems (Berlin: Springer)

\bibitem[{{Hansen} \& {Travis}(1974)}]{Hansen1974eff}
{Hansen}, J.~E., \& {Travis}, L.~D. 1974, \ssr, 16, 527, \dodoi{10.1007/BF00168069}

\bibitem[{{Harrington} {et~al.}(2022){Harrington}, {Himes}, {Cubillos}, {Blecic}, {Rojo}, {Challener}, {Lust}, {Bowman}, {Blumenthal}, {Dobbs-Dixon}, {Foster}, {Foster}, {Green}, {Loredo}, {McIntyre}, {Stemm}, \& {Wright}}]{Harrington2022Bart}
{Harrington}, J., {Himes}, M.~D., {Cubillos}, P.~E., {et~al.} 2022, \psj, 3, 80, \dodoi{10.3847/PSJ/ac3513}

\bibitem[{{Helling}(2023)}]{helling2022cloud}
{Helling}, C. 2023, in {Planetary Systems Now}, ed. L.~M. {Lara} \& D.~{Jewitt} (Singapore: World Scientific)

\bibitem[{{Helling} {et~al.}(2019){Helling}, {Iro}, {Corrales}, {Samra}, {Ohno}, {Alam}, {Steinrueck}, {Lew}, {Molaverdikhani}, {MacDonald}, {Herbort}, {Woitke}, \& {Parmentier}}]{helling2019cloudmap}
{Helling}, C., {Iro}, N., {Corrales}, L., {et~al.} 2019, \aap, 631, A79, \dodoi{10.1051/0004-6361/201935771}

\bibitem[{{Helling} {et~al.}(2023){Helling}, {Samra}, {Lewis}, {Calder}, {Hirst}, {Woitke}, {Baeyens}, {Carone}, {Herbort}, \& {Chubb}}]{Helling2023gridcloud}
{Helling}, C., {Samra}, D., {Lewis}, D., {et~al.} 2023, \aap, 671, A122, \dodoi{10.1051/0004-6361/202243956}

\bibitem[{{Hourdin} {et~al.}(2006){Hourdin}, {Musat}, {Bony}, {Braconnot}, {Codron}, {Dufresne}, {Fairhead}, {Filiberti}, {Friedlingstein}, {Grandpeix}, {Krinner}, {Levan}, {Li}, \& {Lott}}]{Hourdin2006LMDZ}
{Hourdin}, F., {Musat}, I., {Bony}, S., {et~al.} 2006, Climate Dynamics, 27, 787, \dodoi{10.1007/s00382-006-0158-0}

\bibitem[{{Husser} {et~al.}(2013){Husser}, {Wende-von Berg}, {Dreizler}, {Homeier}, {Reiners}, {Barman}, \& {Hauschildt}}]{Husser2013phoenix}
{Husser}, T.~O., {Wende-von Berg}, S., {Dreizler}, S., {et~al.} 2013, \aap, 553, A6, \dodoi{10.1051/0004-6361/201219058}

\bibitem[{Irwin {et~al.}(2008)Irwin, Teanby, De~Kok, Fletcher, Howett, Tsang, Wilson, Calcutt, Nixon, \& Parrish}]{irwin2008nemesis}
Irwin, P., Teanby, N., De~Kok, R., {et~al.} 2008, Journal of Quantitative Spectroscopy and Radiative Transfer, 109, 1136, \dodoi{10.1016/j.jqsrt.2007.11.006}

\bibitem[{{JWST Transiting Exoplanet Community Early Release Science Team}(2022)}]{jwsters2022wasp39b}
{JWST Transiting Exoplanet Community Early Release Science Team}. 2022, \nat, \dodoi{10.1038/s41586-022-05269-w}

\bibitem[{{Karman} {et~al.}(2019){Karman}, {Gordon}, {van der Avoird}, {Baranov}, {Boulet}, {Drouin}, {Groenenboom}, {Gustafsson}, {Hartmann}, {Kurucz}, {Rothman}, {Sun}, {Sung}, {Thalman}, {Tran}, {Wishnow}, {Wordsworth}, {Vigasin}, {Volkamer}, \& {van der Zande}}]{Karman2019hitran}
{Karman}, T., {Gordon}, I.~E., {van der Avoird}, A., {et~al.} 2019, \icarus, 328, 160, \dodoi{10.1016/j.icarus.2019.02.034}

\bibitem[{Kawashima \& Ikoma(2018)}]{kawashima2018theoretical}
Kawashima, Y., \& Ikoma, M. 2018, \apj, 853, 7, \dodoi{10.3847/1538-4357/aaa0c5}

\bibitem[{{Kitzmann} {et~al.}(2020){Kitzmann}, {Heng}, {Oreshenko}, {Grimm}, {Apai}, {Bowler}, {Burgasser}, \& {Marley}}]{Kitzmann2020helios}
{Kitzmann}, D., {Heng}, K., {Oreshenko}, M., {et~al.} 2020, \apj, 890, 174, \dodoi{10.3847/1538-4357/ab6d71}

\bibitem[{Komacek {et~al.}(2019)Komacek, Showman, \& Parmentier}]{Komacek_2019}
Komacek, T.~D., Showman, A.~P., \& Parmentier, V. 2019, \apj, 881, 152, \dodoi{10.3847/1538-4357/ab338b}

\bibitem[{Kreidberg {et~al.}(2014)Kreidberg, Bean, D{\'e}sert, Benneke, Deming, Stevenson, Seager, Berta-Thompson, Seifahrt, \& Homeier}]{kreidberg2014clouds}
Kreidberg, L., Bean, J.~L., D{\'e}sert, J.-M., {et~al.} 2014, \nat, 505, 69, \dodoi{10.1038/nature12888}

\bibitem[{Lavvas {et~al.}(2008)Lavvas, Coustenis, \& Vardavas}]{lavvas2008coupling}
Lavvas, P., Coustenis, A., \& Vardavas, I. 2008, \planss, 56, 67, \dodoi{10.1016/j.pss.2007.05.027}

\bibitem[{{Lee} {et~al.}(2012){Lee}, {Fletcher}, \& {Irwin}}]{Lee2012retrieval}
{Lee}, J.~M., {Fletcher}, L.~N., \& {Irwin}, P.~G.~J. 2012, \mnras, 420, 170, \dodoi{10.1111/j.1365-2966.2011.20013.x}

\bibitem[{Lewis(1969)}]{lewis1969clouds}
Lewis, J.~S. 1969, Icarus, 10, 365, \dodoi{10.1016/0019-1035(69)90091-8}

\bibitem[{Line {et~al.}(2013)Line, Wolf, Zhang, Knutson, Kammer, Ellison, Deroo, Crisp, \& Yung}]{line2013systematic}
Line, M.~R., Wolf, A.~S., Zhang, X., {et~al.} 2013, \apj, 775, 137, \dodoi{10.1088/0004-637X/775/2/137}

\bibitem[{{Lueber} {et~al.}(2022){Lueber}, {Kitzmann}, {Bowler}, {Burgasser}, \& {Heng}}]{lueber2022cloud}
{Lueber}, A., {Kitzmann}, D., {Bowler}, B.~P., {Burgasser}, A.~J., \& {Heng}, K. 2022, \apj, 930, 136, \dodoi{10.3847/1538-4357/ac63b9}

\bibitem[{Lunine {et~al.}(1989)Lunine, Hubbard, Burrows, Wang, \& Garlow}]{lunine1989effect}
Lunine, J.~I., Hubbard, W., Burrows, A., Wang, Y.-P., \& Garlow, K. 1989, \apj, 338, 314, \dodoi{10.1086/167201}

\bibitem[{MacDonald \& Madhusudhan(2017)}]{macdonald2017hd}
MacDonald, R.~J., \& Madhusudhan, N. 2017, \mnras, 469, 1979, \dodoi{10.1093/mnras/stx804}

\bibitem[{Madhusudhan(2019)}]{madhusudhan2019exoplanetary}
Madhusudhan, N. 2019, \araa, 57, 617, \dodoi{10.1146/annurev-astro-081817-051846}

\bibitem[{Madhusudhan {et~al.}(2020)Madhusudhan, Nixon, Welbanks, Piette, \& Booth}]{madhusudhan2020interior}
Madhusudhan, N., Nixon, M.~C., Welbanks, L., Piette, A.~A., \& Booth, R.~A. 2020, \apjl, 891, L7, \dodoi{10.3847/2041-8213/ab7229}

\bibitem[{{Madhusudhan} \& {Seager}(2009)}]{MadhusudhanSeager2009retrieval}
{Madhusudhan}, N., \& {Seager}, S. 2009, \apj, 707, 24, \dodoi{10.1088/0004-637X/707/1/24}

\bibitem[{{Mai} \& {Line}(2019)}]{Mai2019cloudcompare}
{Mai}, C., \& {Line}, M.~R. 2019, \apj, 883, 144, \dodoi{10.3847/1538-4357/ab3e6d}

\bibitem[{Marley {et~al.}(1999)Marley, Gelino, Stephens, Lunine, \& Freedman}]{marley1999reflected}
Marley, M.~S., Gelino, C., Stephens, D., Lunine, J.~I., \& Freedman, R. 1999, \apj, 513, 879, \dodoi{10.1086/306881}

\bibitem[{Min {et~al.}(2020)Min, Ormel, Chubb, Helling, \& Kawashima}]{min2020arcis}
Min, M., Ormel, C.~W., Chubb, K., Helling, C., \& Kawashima, Y. 2020, \aap, 642, A28, \dodoi{10.1051/0004-6361/201937377}

\bibitem[{{Min, M.} {et~al.}(2005){Min, M.}, {Hovenier, J. W.}, \& {de Koter, A.}}]{Min2005DHS}
{Min, M.}, {Hovenier, J. W.}, \& {de Koter, A.} 2005, A\&A, 432, 909, \dodoi{10.1051/0004-6361:20041920}

\bibitem[{{Molli\`ere, P.} {et~al.}(2019){Molli\`ere, P.}, {Wardenier, J. P.}, {van Boekel, R.}, {Henning, Th.}, {Molaverdikhani, K.}, \& {Snellen, I. A. G.}}]{Molli2019DHS}
{Molli\`ere, P.}, {Wardenier, J. P.}, {van Boekel, R.}, {et~al.} 2019, A\&A, 627, A67, \dodoi{10.1051/0004-6361/201935470}

\bibitem[{{Molli\`ere, P.} {et~al.}(2020){Molli\`ere, P.}, {Stolker, T.}, {Lacour, S.}, {Otten, G. P. P. L.}, {Shangguan, J.}, {Charnay, B.}, {Molyarova, T.}, {Nowak, M.}, {Henning, Th.}, {Marleau, G.-D.}, {Semenov, D. A.}, {van Dishoeck, E.}, {Eisenhauer, F.}, {Garcia, P.}, {Garcia Lopez, R.}, {Girard, J. H.}, {Greenbaum, A. Z.}, {Hinkley, S.}, {Kervella, P.}, {Kreidberg, L.}, {Maire, A.-L.}, {Nasedkin, E.}, {Pueyo, L.}, {Snellen, I. A. G.}, {Vigan, A.}, {Wang, J.}, {de Zeeuw, P. T.}, \& {Zurlo, A.}}]{Molli2020petitTRANS}
{Molli\`ere, P.}, {Stolker, T.}, {Lacour, S.}, {et~al.} 2020, A\&A, 640, A131, \dodoi{10.1051/0004-6361/202038325}

\bibitem[{{Nixon} \& {Madhusudhan}(2022)}]{nixon2022aura3d}
{Nixon}, M.~C., \& {Madhusudhan}, N. 2022, \apj, 935, 73, \dodoi{10.3847/1538-4357/ac7c09}

\bibitem[{{Ormel} \& {Min}(2019)}]{ormel2019arcis}
{Ormel}, C.~W., \& {Min}, M. 2019, \aap, 622, A121, \dodoi{10.1051/0004-6361/201833678}

\bibitem[{Parmentier {et~al.}(2013)Parmentier, Showman, \& Lian}]{parmentier2013}
Parmentier, V., Showman, A.~P., \& Lian, Y. 2013, A\&A, 558, A91, \dodoi{10.1051/0004-6361/201321132}

\bibitem[{{Pinhas} {et~al.}(2019){Pinhas}, {Madhusudhan}, {Gandhi}, \& {MacDonald}}]{pinhas2019POSEIDION}
{Pinhas}, A., {Madhusudhan}, N., {Gandhi}, S., \& {MacDonald}, R. 2019, \mnras, 482, 1485, \dodoi{10.1093/mnras/sty2544}

\bibitem[{Polyansky {et~al.}(2018)Polyansky, Kyuberis, Zobov, Tennyson, Yurchenko, \& Lodi}]{Polyansky2018pokazatel}
Polyansky, O.~L., Kyuberis, A.~A., Zobov, N.~F., {et~al.} 2018, \mnras, 480, 2597, \dodoi{10.1093/mnras/sty1877}

\bibitem[{{Robbins-Blanch} {et~al.}(2022){Robbins-Blanch}, {Kataria}, {Batalha}, \& {Adams}}]{robbins-blanch2022picaso}
{Robbins-Blanch}, N., {Kataria}, T., {Batalha}, N.~E., \& {Adams}, D.~J. 2022, \apj, 930, 93, \dodoi{10.3847/1538-4357/ac658c}

\bibitem[{{Rooney} {et~al.}(2022){Rooney}, {Batalha}, {Gao}, \& {Marley}}]{rooney2022fsed}
{Rooney}, C.~M., {Batalha}, N.~E., {Gao}, P., \& {Marley}, M.~S. 2022, \apj, 925, 33, \dodoi{10.3847/1538-4357/ac307a}

\bibitem[{Rosner(2012)}]{rosner2012transport}
Rosner, D.~E. 2012, Transport processes in chemically reacting flow systems (New York: Dover)

\bibitem[{{Roudier} {et~al.}(2021){Roudier}, {Swain}, {Gudipati}, {West}, {Estrela}, \& {Zellem}}]{Roudier2021Diseq}
{Roudier}, G.~M., {Swain}, M.~R., {Gudipati}, M.~S., {et~al.} 2021, \aj, 162, 37, \dodoi{10.3847/1538-3881/abfdad}

\bibitem[{Sing {et~al.}(2016)Sing, Fortney, Nikolov, Wakeford, Kataria, Evans, Aigrain, Ballester, Burrows, Deming, {et~al.}}]{sing2016continuum}
Sing, D.~K., Fortney, J.~J., Nikolov, N., {et~al.} 2016, \nat, 529, 59, \dodoi{10.1038/nature16068}

\bibitem[{Stevenson(2016)}]{Stevenson2016cloudpopulation}
Stevenson, K.~B. 2016, \apjl, 817, L16, \dodoi{10.3847/2041-8205/817/2/l16}

\bibitem[{{Sumlin} {et~al.}(2018){Sumlin}, {Heinson}, \& {Chakrabarty}}]{pymiescatt2018}
{Sumlin}, B.~J., {Heinson}, W.~R., \& {Chakrabarty}, R.~K. 2018, \jqsrt, 205, 127, \dodoi{10.1016/j.jqsrt.2017.10.012}

\bibitem[{Tennyson \& Yurchenko(2012)}]{tennyson2012exomol}
Tennyson, J., \& Yurchenko, S.~N. 2012, \mnras, 425, 21, \dodoi{10.1111/j.1365-2966.2012.21440.x}

\bibitem[{{Tennyson} \& {Yurchenko}(2021)}]{Tennyson2021exomol}
{Tennyson}, J., \& {Yurchenko}, S.~N. 2021, Astronomy and Geophysics, 62, 6.16, \dodoi{10.1093/astrogeo/atab102}

\bibitem[{{Tinetti} {et~al.}(2013){Tinetti}, {Encrenaz}, \& {Coustenis}}]{Tinetti2013aareview}
{Tinetti}, G., {Encrenaz}, T., \& {Coustenis}, A. 2013, \aapr, 21, 63, \dodoi{10.1007/s00159-013-0063-6}

\bibitem[{Tinetti {et~al.}(2018)Tinetti, Drossart, Eccleston, Hartogh, Heske, Leconte, Micela, Ollivier, Pilbratt, Puig, {et~al.}}]{tinetti2018chemical}
Tinetti, G., Drossart, P., Eccleston, P., {et~al.} 2018, Experimental Astronomy, 46, 135, \dodoi{10.1007/s10686-018-9598-x}

\bibitem[{{Tinetti} {et~al.}(2021){Tinetti}, {Eccleston}, {Haswell}, {Lagage}, {Leconte}, {L{\"u}ftinger}, {Micela}, {Min}, {Pilbratt}, {Puig}, {Swain}, {Testi}, {Turrini}, {Vandenbussche}, {Rosa Zapatero Osorio}, {Aret}, {Beaulieu}, {Buchhave}, {Ferus}, {Griffin}, {Guedel}, {Hartogh}, {Machado}, {Malaguti}, {Pall{\'e}}, {Rataj}, {Ray}, {Ribas}, {Szab{\'o}}, {Tan}, {Werner}, {Ratti}, {Scharmberg}, {Salvignol}, {Boudin}, {Halain}, {Haag}, {Crouzet}, {Kohley}, {Symonds}, {Renk}, {Caldwell}, {Abreu}, {Alonso}, {Amiaux}, {Berth{\'e}}, {Bishop}, {Bowles}, {Carmona}, {Coffey}, {Colom{\'e}}, {Crook}, {D{\'e}sjonqueres}, {D{\'\i}az}, {Drummond}, {Focardi}, {G{\'o}mez}, {Holmes}, {Krijger}, {Kovacs}, {Hunt}, {Machado}, {Morgante}, {Ollivier}, {Ottensamer}, {Pace}, {Pagano}, {Pascale}, {Pearson}, {M{\o}ller Pedersen}, {Pniel}, {Roose}, {Savini}, {Stamper}, {Szirovicza}, {Szoke}, {Tosh}, {Vilardell}, {Barstow}, {Borsato}, {Casewell}, {Changeat}, {Charnay}, {Civi{\v{s}}}, {Coud{\'e} du Foresto}, {Coustenis}, {Cowan},
  {Danielski}, {Demangeon}, {Drossart}, {Edwards}, {Gilli}, {Encrenaz}, {Kiss}, {Kokori}, {Ikoma}, {Morales}, {Mendon{\c{c}}a}, {Moneti}, {Mugnai}, {Garc{\'\i}a Mu{\~n}oz}, {Helled}, {Kama}, {Miguel}, {Nikolaou}, {Pagano}, {Panic}, {Rengel}, {Rickman}, {Rocchetto}, {Sarkar}, {Selsis}, {Tennyson}, {Tsiaras}, {Venot}, {Vida}, {Waldmann}, {Yurchenko}, {Szab{\'o}}, {Zellem}, {Al-Refaie}, {Perez Alvarez}, {Anisman}, {Arhancet}, {Ateca}, {Baeyens}, {Barnes}, {Bell}, {Benatti}, {Biazzo}, {B{\l}{\k{e}}cka}, {Bonomo}, {Bosch}, {Bossini}, {Bourgalais}, {Brienza}, {Brucalassi}, {Bruno}, {Caines}, {Calcutt}, {Campante}, {Canestrari}, {Cann}, {Casali}, {Casas}, {Cassone}, {Cara}, {Carmona}, {Carone}, {Carrasco}, {Changeat}, {Chioetto}, {Cortecchia}, {Czupalla}, {Chubb}, {Ciaravella}, {Claret}, {Claudi}, {Codella}, {Garcia Comas}, {Cracchiolo}, {Cubillos}, {Da Peppo}, {Decin}, {Dejabrun}, {Delgado-Mena}, {Di Giorgio}, {Diolaiti}, {Dorn}, {Doublier}, {Doumayrou}, {Dransfield}, {Dumaye}, {Dunford}, {Jimenez Escobar}, {Van
  Eylen}, {Farina}, {Fedele}, {Fern{\'a}ndez}, {Fleury}, {Fonte}, {Fontignie}, {Fossati}, {Funke}, {Galy}, {Garai}, {Garc{\'\i}a}, {Garc{\'\i}a-Rigo}, {Garufi}, {Germano Sacco}, {Giacobbe}, {G{\'o}mez}, {Gonzalez}, {Gonzalez-Galindo}, {Grassi}, {Griffith}, {Guarcello}, {Goujon}, {Gressier}, {Grzegorczyk}, {Guillot}, {Guilluy}, {Hargrave}, {Hellin}, {Herrero}, {Hills}, {Horeau}, {Ito}, {Jessen}, {Kabath}, {K{\'a}lm{\'a}n}, {Kawashima}, {Kimura}, {Kn{\'\i}{\v{z}}ek}, {Kreidberg}, {Kruid}, {Kruijssen}, {Kubel{\'\i}k}, {Lara}, {Lebonnois}, {Lee}, {Lefevre}, {Lichtenberg}, {Locci}, {Lombini}, {Sanchez Lopez}, {Lorenzani}, {MacDonald}, {Magrini}, {Maldonado}, {Marcq}, {Migliorini}, {Modirrousta-Galian}, {Molaverdikhani}, {Molinari}, {Molli{\`e}re}, {Moreau}, {Morello}, {Morinaud}, {Morvan}, {Moses}, {Mouzali}, {Nakhjiri}, {Naponiello}, {Narita}, {Nascimbeni}, {Nikolaou}, {Noce}, {Oliva}, {Palladino}, {Papageorgiou}, {Parmentier}, {Peres}, {P{\'e}rez}, {Perez-Hoyos}, {Perger}, {Cecchi Pestellini}, {Petralia},
  {Philippon}, {Piccialli}, {Pignatari}, {Piotto}, {Podio}, {Polenta}, {Preti}, {Pribulla}, {Lopez Puertas}, {Rainer}, {Reess}, {Rimmer}, {Robert}, {Rosich}, {Rossi}, {Rust}, {Saleh}, {Sanna}, {Schisano}, {Schreiber}, {Schwartz}, {Scippa}, {Seli}, {Shibata}, {Simpson}, {Shorttle}, {Skaf}, {Skup}, {Sobiecki}, {Sousa}, {Sozzetti}, {{\v{S}}poner}, {Steiger}, {Tanga}, {Tackley}, {Taylor}, {Tecza}, {Terenzi}, {Tremblin}, {Tozzi}, {Triaud}, {Trompet}, {Tsai}, {Tsantaki}, {Valencia}, {Carine Vandaele}, {Van der Swaelmen}, {Vardan}, {Vasisht}, {Vazan}, {Del Vecchio}, {Waltham}, {Wawer}, {Widemann}, {Wolkenberg}, {Hou Yip}, {Yung}, {Zilinskas}, {Zingales}, \& {Zuppella}}]{tinetti_ariel2}
{Tinetti}, G., {Eccleston}, P., {Haswell}, C., {et~al.} 2021, Tech. rep., ESA.
\newblock \url{https://www.cosmos.esa.int/documents/1783156/3267291/Ariel_RedBook_Nov2020.pdf}

\bibitem[{{Toon} {et~al.}(1979){Toon}, {Turco}, {Hamill}, {Kiang}, \& {Whitten}}]{Toon1979CARMA}
{Toon}, O.~B., {Turco}, R.~P., {Hamill}, P., {Kiang}, C.~S., \& {Whitten}, R.~C. 1979, Journal of Atmospheric Sciences, 36, 718, \dodoi{10.1175/1520-0469(1979)036<0718:AODMDA>2.0.CO;2}

\bibitem[{Tremblin {et~al.}(2015)Tremblin, Amundsen, Mourier, Baraffe, Chabrier, Drummond, Homeier, \& Venot}]{Tremblin2015ATMO}
Tremblin, P., Amundsen, D.~S., Mourier, P., {et~al.} 2015, \apjl, 804, L17, \dodoi{10.1088/2041-8205/804/1/l17}

\bibitem[{{Tsai} {et~al.}(2021){Tsai}, {Innes}, {Lichtenberg}, {Taylor}, {Malik}, {Chubb}, \& {Pierrehumbert}}]{tsai2021infer}
{Tsai}, S.-M., {Innes}, H., {Lichtenberg}, T., {et~al.} 2021, \apjl, 922, L27, \dodoi{10.3847/2041-8213/ac399a}

\bibitem[{Tsiaras {et~al.}(2019)Tsiaras, Waldmann, Tinetti, Tennyson, \& Yurchenko}]{tsiaras2019water}
Tsiaras, A., Waldmann, I.~P., Tinetti, G., Tennyson, J., \& Yurchenko, S.~N. 2019, Nature Astronomy, 3, 1086, \dodoi{10.1038/s41550-019-0878-9}

\bibitem[{{Tsiaras} {et~al.}(2018){Tsiaras}, {Waldmann}, {Zingales}, {Rocchetto}, {Morello}, {Damiano}, {Karpouzas}, {Tinetti}, {McKemmish}, {Tennyson}, \& {Yurchenko}}]{Tsiaras2018population}
{Tsiaras}, A., {Waldmann}, I.~P., {Zingales}, T., {et~al.} 2018, \aj, 155, 156, \dodoi{10.3847/1538-3881/aaaf75}

\bibitem[{{Turco} {et~al.}(1979){Turco}, {Hamill}, {Toon}, {Whitten}, \& {Kiang}}]{Turco1979CARMA}
{Turco}, R.~P., {Hamill}, P., {Toon}, O.~B., {Whitten}, R.~C., \& {Kiang}, C.~S. 1979, Journal of Atmospheric Sciences, 36, 699, \dodoi{10.1175/1520-0469(1979)036<0699:AODMDA>2.0.CO;2}

\bibitem[{Venot {et~al.}(2020)Venot, Parmentier, Blecic, Cubillos, Waldmann, Changeat, Moses, Tremblin, Crouzet, Gao, {et~al.}}]{venot2020global}
Venot, O., Parmentier, V., Blecic, J., {et~al.} 2020, \apj, 890, 176, \dodoi{10.3847/1538-4357/ab6a94}

\bibitem[{{Wakeford} {et~al.}(2018){Wakeford}, {Sing}, {Deming}, {Lewis}, {Goyal}, {Wilson}, {Barstow}, {Kataria}, {Drummond}, {Evans}, {Carter}, {Nikolov}, {Knutson}, {Ballester}, \& {Mandell}}]{wakeford2018cloudretrieval}
{Wakeford}, H.~R., {Sing}, D.~K., {Deming}, D., {et~al.} 2018, \aj, 155, 29, \dodoi{10.3847/1538-3881/aa9e4e}

\bibitem[{Wang {et~al.}(2022)Wang, Fujii, \& He}]{wang2022nongray}
Wang, F., Fujii, Y., \& He, J. 2022, \apj, 931, 48, \dodoi{10.3847/1538-4357/ac67e5}

\bibitem[{{Warren} \& {Brandt}(2008)}]{warren2008refractive}
{Warren}, S.~G., \& {Brandt}, R.~E. 2008, Journal of Geophysical Research (Atmospheres), 113, D14220, \dodoi{10.1029/2007JD009744}

\bibitem[{{Welbanks} \& {Madhusudhan}(2021)}]{welbanks2021aurora}
{Welbanks}, L., \& {Madhusudhan}, N. 2021, \apj, 913, 114, \dodoi{10.3847/1538-4357/abee94}

\bibitem[{{Welbanks} {et~al.}(2019){Welbanks}, {Madhusudhan}, {Allard}, {Hubeny}, {Spiegelman}, \& {Leininger}}]{Welbanks2019Trends}
{Welbanks}, L., {Madhusudhan}, N., {Allard}, N.~F., {et~al.} 2019, \apjl, 887, L20, \dodoi{10.3847/2041-8213/ab5a89}

\bibitem[{{Windsor} {et~al.}(2023){Windsor}, {Robinson}, {Kopparapu}, {Young}, {Trilling}, \& {LLama}}]{Windsor2023cloud}
{Windsor}, J.~D., {Robinson}, T.~D., {Kopparapu}, R.~k., {et~al.} 2023, \psj, 4, 94, \dodoi{10.3847/PSJ/acbf2d}

\bibitem[{{Xuan} {et~al.}(2022){Xuan}, {Wang}, {Ruffio}, {Knutson}, {Mawet}, {Molli{\`e}re}, {Kolecki}, {Vigan}, {Mukherjee}, {Wallack}, {Wang}, {Baker}, {Bartos}, {Blake}, {Bond}, {Bryan}, {Calvin}, {Cetre}, {Chun}, {Delorme}, {Doppmann}, {Echeverri}, {Finnerty}, {Fitzgerald}, {Horstman}, {Inglis}, {Jovanovic}, {L{\'o}pez}, {Martin}, {Morris}, {Pezzato}, {Ragland}, {Ren}, {Ruane}, {Sappey}, {Schofield}, {Skemer}, {Venenciano}, {Wallace}, \& {Wizinowich}}]{xuan2022cloudbrown}
{Xuan}, J.~W., {Wang}, J., {Ruffio}, J.-B., {et~al.} 2022, \apj, 937, 54, \dodoi{10.3847/1538-4357/ac8673}

\bibitem[{Yu {et~al.}(2021)Yu, Moses, Fortney, \& Zhang}]{yu2021identify}
Yu, X., Moses, J.~I., Fortney, J.~J., \& Zhang, X. 2021, \apj, 914, 38, \dodoi{10.3847/1538-4357/abfdc7}

\bibitem[{{Zhang} {et~al.}(2020){Zhang}, {Chachan}, {Kempton}, {Knutson}, \& {Chang}}]{Zhang2020platon}
{Zhang}, M., {Chachan}, Y., {Kempton}, E. M.~R., {Knutson}, H.~A., \& {Chang}, W.~H. 2020, \apj, 899, 27, \dodoi{10.3847/1538-4357/aba1e6}

\bibitem[{Zhang \& Showman(2018)}]{Zhang_2018}
Zhang, X., \& Showman, A.~P. 2018, \apj, 866, 1, \dodoi{10.3847/1538-4357/aada85}

\end{thebibliography}

\bibliographystyle{aasjournal}



\end{document}